\documentclass[reprint,twocolumn,superscriptaddress,showpacs,nofootinbib,notitlepage,preprintnumbers,aps,prd,floatfix]{revtex4-2}
\usepackage[utf8]{inputenc}
\usepackage{graphicx}
\usepackage{latexsym,amsmath,amssymb,lmodern,float,url}
\usepackage{natbib}
\usepackage{color}
\usepackage{microtype}
\usepackage{bbold}
\usepackage{slashed}
\usepackage{multirow}
\usepackage{tikz}
\usepackage{xcolor}
\usepackage{mathrsfs}  
\usetikzlibrary{shapes}

\usepackage[colorlinks=true,backref=false, linktocpage=true,
citecolor=blue,urlcolor=blue,linkcolor=blue,pdfpagemode=UseOutlines]{hyperref}

\hypersetup{%
 bookmarksnumbered=true,
 pdftitle = {},
 pdfsubject = {},
 pdfauthor = {},
 pdfkeywords = {}
}

\let\Re\undefined
\let\Im\undefined
\newcommand{\zn}{\mathbb{Z}_{\rm n}}
\newcommand{\bi}{\mathbb{BI}}
\newcommand{\ds}{\mathbb{V}}
\newcommand{\suep}{S[u,\epsilon]}
\newcommand{\upfun }{}

\DeclareMathOperator{\Tr}{Tr}
\DeclareMathOperator{\tr}{Tr}
\DeclareMathOperator{\Re}{Re}
\DeclareMathOperator{\Im}{Im}
\newcommand \vev [1] {\langle{#1}\rangle}

\def \Im {\mathop{\rm Im}\nolimits}

\definecolor{mycolor}{RGB}{0,255,0}
\newcommand{\redsq}{\raisebox{0.5pt}{\tikz{\node[fill,scale=0.4,regular polygon, regular polygon sides=4,fill=red](){};}}}
\newcommand{\gcir}{\raisebox{0.5pt}{\tikz{\node[fill,scale=0.4,circle,fill=mycolor](){};}}}
\newcommand{\bltri}{\raisebox{0.7pt}{\tikz{\node[fill,scale=0.3,regular polygon, regular polygon sides=3,fill=blue!10!blue,rotate=0](){};}}}

\begin{document}
\preprint{FERMILAB-PUB-20-198-T, SI-HEP-2020-25}
\title{Gluon Field Digitization via Group Space Decimation for Quantum Computers}
\author{Yao Ji}
\email{yao.ji@uni-siegen.de}
\affiliation{Theoretische Physik 1, Naturwissenschaftlich-Technische Fakult\"at, Universit\"at Siegen, D-57068 Siegen, Germany}  
\affiliation{Institut f\"ur Theoretische Physik, Universit\"at Regensburg, D-93040 Regensburg, Germany}
\author{Henry Lamm}
\email{hlamm@fnal.gov}
\affiliation{Fermi National Accelerator Laboratory, Batavia, Illinois, 60510, USA}
\author{Shuchen Zhu}
\email{sz424@georgetown.edu} 
\affiliation{Department of Computer Science, Georgetown University, Washington, DC 20057, USA}
\date{\today}
\collaboration{NuQS Collaboration}
%%%%%%%%%%%%%%%%%%%%%%%%%%%%%%%%%%%%%%%%%%%%%%%%%%%%%%%
\begin{abstract}
Efficient digitization is required for quantum simulations of gauge theories. Schemes based on discrete subgroups use fewer qubits at the cost of systematic errors. We systematize this approach by deriving a single plaquette action for approximating general continuous gauge groups through integrating out field fluctuations. This provides insight into the effectiveness of these approximations, and how they could be improved. We accompany the scheme by simulations of pure gauge over the largest discrete subgroup of $SU(3)$ up to the third order. 
\end{abstract}
%%%%%%%%%%%%%%%%%%%%%%%%%%%%%%%%%%%%%%%%%%%%%%%%%%
\maketitle
\section{Introduction}
Large-scale quantum computers can simulate nonperturbative quantum field theories which are intractable classically~\cite{Feynman:1981tf}. Alas, Noisy Intermediate-Scale Quantum (NISQ) era systems will be limited both in qubits and circuit depths. Whether any gauge theory simulations in this period are possible depends upon efficient formulations. The situation is similar to the early days of lattice field theory when computer memory was limited and the cost of storing $SU(3)$ elements was prohibitive. 

For fermionic fields, relatively efficient mappings to quantum registers are known~\cite{Jordan:1928wi,Verstraete:2005pn,Zohar:2018cwb,2016PhRvA..94c0301W} evidenced by most existing quantum calculations being fermionic~\cite{Martinez:2016yna,Klco:2018kyo,Lamm:2018siq,Shehab:2019gfn}. The bosonic nature of gauge fields preclude exact mappings, but many proposals exist with different costs~\cite{Zohar:2013zla,Hackett:2018cel,Macridin:2018gdw,Yeter-Aydeniz:2018mix,Klco:2018zqz,Bazavov:2015kka,Zhang:2018ufj,Unmuth-Yockey:2018xak,Unmuth-Yockey:2018ugm,Zache:2018jbt,Raychowdhury:2018osk,Kaplan:2018vnj,Stryker:2018efp,Alexandru:2019ozf,Chandrasekharan:1996ih,Schlittgen:2000xg,Brower:1997ha,Klco:2019evd,Alexandru:2019nsa}. Digitizing reduces symmetries -- either explicitly or through finite-truncations~\cite{Zohar:2013zla}. These breakings mean \textit{a priori} the original model may not be recovered in the continuum limit~\cite{Hasenfratz:2001iz,Caracciolo:2001jd,Hasenfratz:2000hd,PhysRevE.57.111,PhysRevE.94.022134,article}. Further, choices of digitization may limit the use of classical resources for Euclidean simulations or state preparation~\cite{Harmalkar:2020mpd}. In summary, the understanding of resource costs, systematic errors, and the continuum limit for these proposals is poorly known today. 

In this work, we systematize the proposal of replacing continuous gauge groups $G$ by their discrete subgroups $H$~\cite{Hackett:2018cel,Alexandru:2019nsa} by deriving lattice actions using the group space decimation procedure of~\cite{Flyvbjerg:1984dj,Flyvbjerg:1984ji}. After deriving the general third order action, we will investigate the behaviour of discretizing three distinct gauge groups $U(1)$, $SU(2)$, and $SU(3)$.  
We begin by reviewing the discrete subgroup approximation in Sec.~\ref{sec:ds}.  In Sec.~\ref{sec:gsd} we discuss the general aspects of the group space decimation procedure. Following that, in Sec.~\ref{sec:obo} we derive the decimated action up to 3rd order. Numerical results for $\ds$ using the decimated actions are presented in Sec.~\ref{sec:num}. Sec.~\ref{sec:cont} studies the continuous group limit, and we conclude in Sec.~\ref{sec:concl}.
\section{Discrete Subgroups}
\label{sec:ds}
Approximating gauge theories by replacing $G\rightarrow H$ was explored in the early days of Euclidean lattice field theory. The viability of the $\zn$ subgroups replacing $U(1)$ were studied in~\cite{Creutz:1979zg,Creutz:1982dn}. 
Further studies of the crystal-like discrete subgroups of $SU(N)$ were performed~\cite{Bhanot:1981xp,Petcher:1980cq,Bhanot:1981pj}, including with fermions~\cite{Weingarten:1980hx,Weingarten:1981jy}.
These studies met with mixed success depending on the group and action tested. 

The fundamental issue of group discretization can be understood by considering the Wilson gauge action \begin{equation}
\label{eq:wils}
S[U]=-\sum_p \frac{\beta}{N}\Re\tr (U_p) \,,
\end{equation}
where $U_p$ indicates a plaquette of continuous group gauge links $U$ (for discrete groups, $u_p$ denotes plaquettes and $u$ denotes links). 
As $\beta\rightarrow \infty$, links near the group identity $\mathbb{1}$ dominate, i.e. $U\approx\mathbb{1}+\varepsilon$, where $\varepsilon$ can be arbitrarily small. Therefore the gap $\Delta S = S[\mathbb{1}+\varepsilon]-S[\mathbb{1}]$ goes to zero smoothly.
For discrete groups, $\varepsilon$ has a minimum given by the nearest elements $\mathcal{N}$ to $\mathbb{1}$, and thus $\Delta S=S[\mathcal{N}]-S[\mathbb{1}]>0$. 
This strongly suggests a phase transition at some critical $\beta_f=c/\Delta S$, where $c\approx\mathcal{O}(1)$ depends on spacetime dimensionality, gauge group, and entropy. 
For $U(1)\rightarrow \zn$ in 4d, $
\beta_f=\frac{0.78}{1-\cos(2\pi/n)}$~\cite{Petcher:1980cq}. 
Above $\beta_f$, all field configurations but $u=\mathbb{1}$ are exponentially suppressed. Thus, $H$ fails to approximate $G$ for $\beta>\beta_f$. 
Another way to understand this behavior follows~\cite{Fradkin:1978dv}, where some discrete theories are shown equivalent to continuous groups coupled to a Higgs field. The Higgs mechanism introduces a new phase missing from the continuous gauge theory when $\beta\rightarrow\infty$. 

Both arguments suggest $H$ be viewed as an \emph{effective field theory} for $G$ with a UV-cutoff at $\Lambda_f$. Provided the typical separation of scales of physics $m_{IR}\ll \Lambda_f$, the approximation could be reliable up to $\mathcal{O}(m_{IR}/\Lambda_f)$ effects. 

In lattice calculations, one replaces $\Lambda_f$ by a fixed lattice spacing $a=a(\beta)$ which shrinks as $\beta\rightarrow\infty$ for asymptotically free theories.
To control errors when extrapolating to $a\rightarrow0$, one should simulate in the \emph{scaling regime} of $a\ll m_{IR}^{-1}$. We denote the onset of the scaling regime by $a_s,$ and $\beta_s$. For $a_f(\beta_f)\sim \Lambda_f^{-1}$,  errors from the discrete group approximation would be small if $a$ can be reduced such that $m_{IR}^{-1}\gg a\gtrsim a_f$ i.e. $\beta_s\leq\beta_f$.

In the case of $U(1)$ with $\beta_s= 1$, $\mathbb{Z}_{n>5}$ satisfies $\beta_f>\beta_s$. 
For non-Abelian groups, only a finite set of crystal-like subgroups exist. $SU(2)$ has three: the binary tetrahedral $\mathbb{BT}$, the binary octahedral $\mathbb{BO}$, and the binary icosahedral $\bi$. 
While $\mathbb{BT}$ has $\beta_f=2.24(8)$, $\mathbb{BO}$ and $\bi$ have $\beta_f=3.26(8)$ and $\beta_f=5.82(8)$ respectively~\cite{Alexandru:2019nsa}, above $\beta_s=2.2$. Hence, $\mathbb{BO}$ and $\bi$ appear useful for $SU(2)$.

For the important case of $SU(3)$ with $\beta_s=6$, there are five crystal-like subgroups with the Valentiner group $\ds$ with 1080 elements\footnote{This name is most common in the mathematical literature~\cite{valentiner1889endelige,crass1999solving}. It has also referred to as $S(1080)$~\cite{Bhanot:1981pj,Flyvbjerg:1984dj,Flyvbjerg:1984ji,Alexandru:2019nsa} or $\Sigma_{3\times360}$~\cite{Hagedorn:2013nra}.}. For all subgroups, $\beta_f < \beta_s$ , with $\ds$ having $\beta_f=3.935(5)$~\cite{Alexandru:2019nsa} and thus appear inadequate. Other work~\cite{Lisboa:1982jj} has shown that extending to a subset with the midpoints between elements of $\ds$ raises $\beta_f\approx7$. However this require more qubits and -- potentially more worrisome-- sacrifices gauge symmetry completely which is dangerous on quantum computers~\cite{Stryker:2018efp,Halimeh:2019svu,1797835}.

To decrease $a_f$, adding additional terms to Eq.~(\ref{eq:wils}) was attempted ~\cite{Edgar:1981dr,Bhanot:1981pj,Creutz:1982dn,Fukugita:1982kk,Horn:1982ef,Flyvbjerg:1984dj,Flyvbjerg:1984ji,Ayala:1989it,Alexandru:2019nsa}, although only in~\cite{Bhanot:1981pj,Alexandru:2019nsa} were Monte Carlo calculations undertaken for $SU(3)$. Two reasons suggest this would help. First, additional terms which have a continuum limit $\propto \Re\Tr F_{\mu\nu}F^{\mu\nu}$, but take different values on the element of $H$ (e.g. $|\Tr (u_p)|^2-1$), change $\Delta S$ and thus $a_f$. Second, new terms can reduce finite-$a$ errors as in Symanzik improvement.

The term usually added was the adjoint trace, giving 
\begin{equation}
\label{eq:actadj}
 S[u]=-\sum_p\left(\frac{{\beta}_{\{1\}}}{3}\Re\tr (u_p) +\frac{{\beta}_{\{1,-1\}}}{8}|\tr (u_p)|^2\right),
\end{equation}
where $u_p \in \ds$, and the first term is normalized so for $\beta_{\{1,-1\}}=0$, the $S[u]$ matches the Wilson action (with $\beta_{\{1\}}=\beta$). In these works, no relationship was assumed between $\beta_{\{1\}}$ and $\beta_{\{1,-1\}}$. That Eq.~\eqref{eq:actadj} improves the viability of $\ds$ over the Eq.~\eqref{eq:wils} will be shown in~\cite{lammtocome}.  For a different action, 
\begin{equation}\label{eq:action-mod}
 S[u]=-\sum_p \left(\frac{\beta_0}{3}\Re\Tr (u_p) +\beta_1\Re\Tr(u_p^2)\right) \,,
\end{equation}
smaller values of $a_f$ were demonstrated in~\cite{Alexandru:2019nsa}.

With these actions, the dimensionless product $T_c\sqrt{t_0}$ of the pseudocritical temperature and the Wilson flow parameter were found to agree in the continuum with $SU(3)$, allowing one to set the scale of those calculations. $a>0.08$ fm was achieved without the effects of $a_f$ being seen. This suggest that $\ds$ can reproduce $SU(3)$ in the scaling region with a modified action, such that practical quantum computations of $SU(3)$ could be performed. While promising, the choice of new terms was ad-hoc and left unclear how to systematically improve or analyze effectiveness. In the next section, we systematically derive lattice actions for $H$, discovering that the terms added in these two actions are in fact the first terms generated. 

\section{Group Space Decimation}
\label{sec:gsd}
Our ultimate goal is to approximate the path integral of group $G$ faithfully by a discrete subgroup $H$ by replacing the integration over $G$ by a summation over $H$. Group space decimation can be understood in analogy to Wilsonian renormalization, where we integrate out continuous field fluctuations instead of UV modes. The typical method used with discrete subgroup approximations is to replace the gauge links $U\in G$ by $u\in H$ such that the action $S[U]\rightarrow S[u]$. This corresponds to simply regularizing a field theory. For strong coupling, this appears sufficient. 
As $\beta\rightarrow \infty$, correlations between gauge links increase and the average field fluctuation becomes smaller. 
When the average field fluctuations decrease below the distance between $\mathbb{1}$ and $\mathcal{N}$ of the discrete group, freeze-out occurs and the approximation breaks down--similar to probing a regulated theory too close to the cutoff.
Therefore, improving this approximation and understand the systematics can be done by considering these discarded continuous field fluctuations. To do this, instead of performing the replacement $U\rightarrow u$, we will integrate out the continuous fluctuations, following the decimation formalism developed by Flyvbjerg~\cite{Flyvbjerg:1984ji,Flyvbjerg:1984dj}. He derived the second order decimated action for $U(1)$, $SU(2)$, and $SU(3)$. An important general feature of the decimated action though is missing from this second order action -- while new terms are generated at each order, until third order no coefficient of an existing term is modified. One expects such terms are critical to understanding deviations from the continuous group and therefore we compute them in Sec.~\ref{sec:obo}.

\begin{figure}
  \centering
  \includegraphics[width=0.5\linewidth]{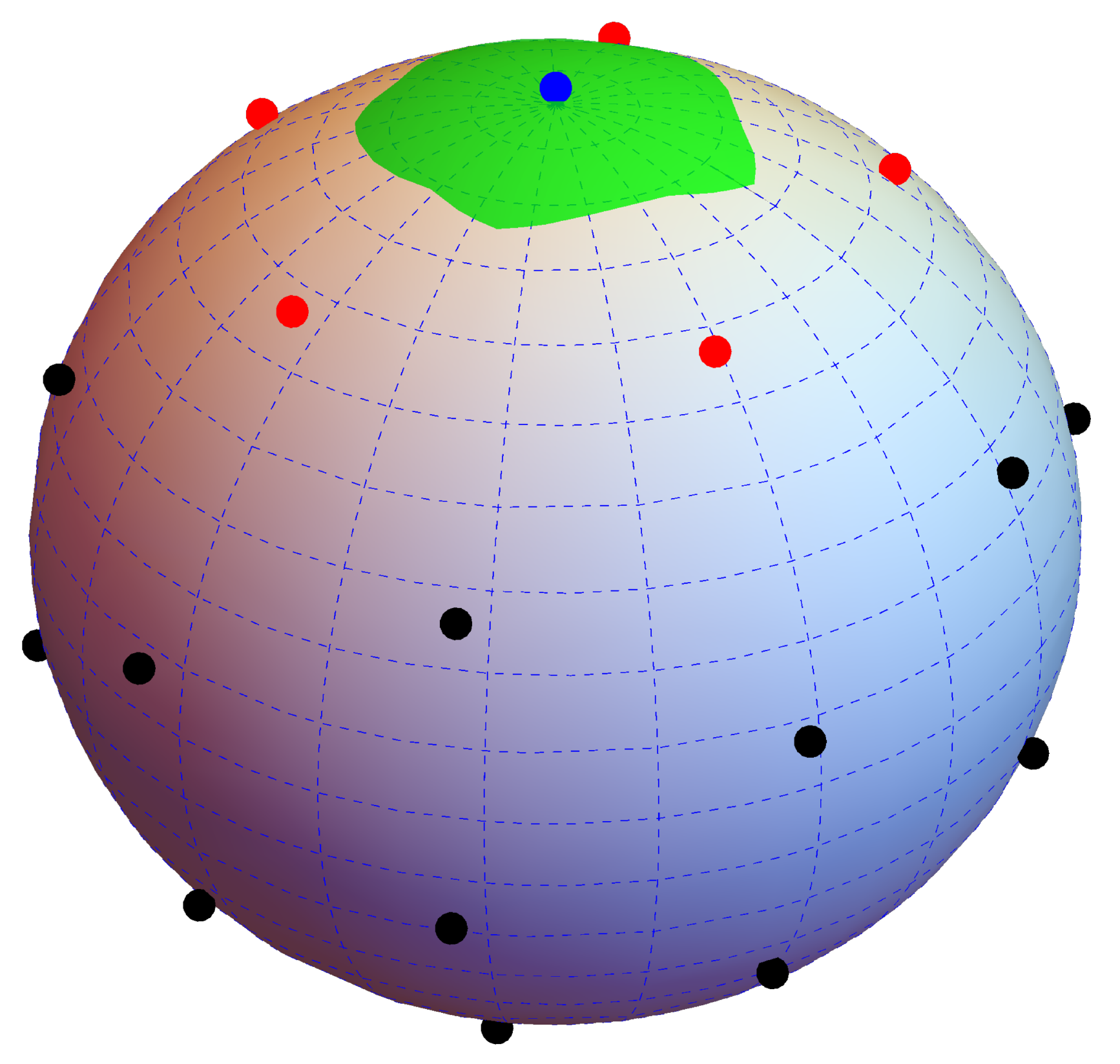}
  \caption{A schematic demonstration of $\Omega$ (in green) of $G$ (a sphere) around $\mathbb{1}$ (blue point) of the discrete group (shown as points). $\mathcal{N}$ for $H$ are given by red points. We have applied the $S^2$ metric to obtain the $\Omega$. In groups representable in two dimensions, this region resembles a polygon while in higher dimensions, it becomes a polytope.} 
  \label{fig:sphere}
\end{figure}

 It is natural to associate every subgroup element $u\in H$ with an unique set, or region, $\Omega_u$ containing all closest continuous group elements $U\in G$:
\begin{align}
\Omega_u\equiv\left\{U\in G\big | d(U,u)<d(U,u'), \forall u'\in H\setminus{\{u\}}\right\}\, ,
\end{align}
where the distance is defined as $d^2(U,u)=\Tr\big((U-u)^\dag (U-u)\big)$. 
By such a definition. the continuous group is fully covered, i.e., $G=\cup_{u\in H}\Omega_u$ and a graphical demonstration of $\Omega\equiv\Omega_{\mathbb{1}}$ can be found in Fig.~\ref{fig:sphere}.
Note that for any $U\in G$, there exist a unique  $u\in H$ and $\epsilon\in \Omega$ such that $U=u\epsilon$, where we may treat $\epsilon$ as the error of $u$ approximating $U$.
In this way, without approximation, the Euclidean path integral integrating over $G$ can be written as a summation over $H$ and integration over $\epsilon\in\Omega$:
\begin{align}
  Z=\int_G DU \,e^{-S[U]} = \sum_{u\in H}\int_{\Omega}D\epsilon\, e^{-S[u,\epsilon]}\,,\label{eq:path integral}
\end{align}
where $Z$ is a functional integral over all gauge links $U$ on the lattice, or equivalently a functional integral over $\epsilon$ and a functional sum over $u$. In this expression, $S[u,\epsilon]=S[U]$ is defined by replacing each gauge link $U$ by $u\epsilon$. 

We then expand the exponential in the path integral and integrate over $\epsilon$ producing a moment expansion
\begin{align}
  Z &=\sum_{u\in H}\int_{\Omega}D\epsilon\, \left(1-\beta \suep +\frac{\beta^2}{2!}\suep^2+\cdots \right)\nonumber\\
  &=\sum_{u\in H}\left(1-\beta\vev{\suep}+ \frac{\beta^2}{2!}\vev{\suep^2}+\cdots \right) \,,\label{eq:z-natural-expansion}
\end{align}
where we have introduced the notation $\langle f\rangle=\int_{\Omega} D\epsilon\, f$ with normalization $\int_{\Omega} D\epsilon=1$.
What we are really after is an expansion for the action $S[U]$, writing $Z$ in terms of a cumulant expansion
\begin{align}
  Z &= \sum_{u\in H}\exp{\left(-\sum_{n=1}^\infty\frac{\beta^n}{n!}\mathscr{S}_n[u]\right)}\,, \label{eq:z-effective}
\end{align}
allows us to match Eq.~\eqref{eq:z-natural-expansion} with \eqref{eq:z-effective} to obtain an effective action.  In this way, after integrating over $\epsilon$, the contributions to the action depend only on the discrete group gauge link $u$ %i.e. $\vev{\suep^n}_c\equiv \mathscr{S}_n[u]$ 
 and the effective action can be defined as 
\begin{align}
    S[u]\equiv\sum_n \frac{\beta^n}{n!}\mathscr{S}_n[u].
\end{align}
Up to $\mathcal{O}(\beta^3)$ one has
\begin{align}
  &\mathscr{S}_1[u]=\vev{\suep}\, ,\label{eq:first_order}\\
  &\mathscr{S}_2[u]=-\vev{\suep^2}+\vev{\suep}^2\, ,\label{eq:second_order}\\
  &\mathscr{S}_3[u]=\vev{\suep^3}-3\vev{\suep}\vev{\suep^2}+2\vev{\suep}^3\,.\label{eq:connected2}
\end{align}
One may worry about poor convergence in the region of interest $\beta\geq\beta_s\geq 1$ . As will be discussed more thoroughly in Sec.~\ref{sec:cont}, $\beta^n$ terms are suppressed by powers of the average field fluctuation. Thus, the size of the discrete group, which determines the size of field fluctuations integrated out, also determines the series convergence.

Starting with the second order terms computed in Refs.~\cite{Flyvbjerg:1984dj,Flyvbjerg:1984ji}, the decimated action generates multi-plaquette contributions. Their inclusion in quantum simulations brings substantial non-locality which requires high qubit connectivity and increases circuit depth. Luckily these contributions will be shown to be small in Sec.~\ref{sec:obo}.  

In the following section, we will calculate Eq.~\eqref{eq:first_order} to \eqref{eq:connected2} in terms of linear combination of the group characters starting from the Wilson action of Eq.~(\ref{eq:wils}):
\begin{align}
  S[U]\equiv -\sum_p\frac{\beta}{N}\Re \tr(U_p)=-\sum_p\frac{\beta}{N}\Re \chi_{\{1\}} \,. \label{eq:action-definition}
\end{align}
Here we introduced $\chi_r$, the character of the group representation\footnote{Through out this work we suppress the argument of $\chi_r$, but it can only be $U_p$ or $u_p$ and context makes it clear which is meant.} $r$. This is the natural basis for the decimated action. All characters required for our $\mathcal{\beta}^3$ calculation are in Table~\ref{tab:vrnum}. In the interest of deriving a decimated action for general gauge groups, we have chosen a nonstandard basis for $U(1)$ and $SU(2)$.  This allows for one general scheme for $U(N)$ and $SU(N)$ groups. This basis is not linearly independent and relations between representations exist. This dependence is typically used to write $U(1)$ and $SU(2)$ in reduced sets of representations.  We have collated relations between the over complete basis in Appendix~\ref{apx:groups}.
 
 In deriving the decimated action, integrating out the field fluctuations require us to reduce expressions of the form $\vev{\epsilon_{i_1j_1}\cdots \epsilon_{i_nj_n}}$. To simplify these, we use an identity derived in~\cite{Creutz:1978ub} for $SU(N)$ and $U(N)$ groups for any integer $n\leq N$.  The necessary relations for  $n\leq3$ are found in Appendix~\ref{apx:eident}. From these identites, we are left with expectation values of $\chi_r$ over $\Omega$

\begin{equation}
\label{eq:v_r_definition}
 V_r\equiv\frac{1}{d_r}\langle\Re \chi_r\rangle \,,
\end{equation}
where $d_r$ is the dimension of representation $r$.

 For $U(1)\rightarrow \zn$, there is only one representation at each order of the cumulant expansion, $V_{\{h\}}=\langle \epsilon^h\rangle$.  These terms can be computed analytically by a change of variables $\epsilon=e^{i\phi}$~\cite{Flyvbjerg:1984ji}:
\begin{align}
 V_{\{h\}} &=\frac{1}{V_{0}}\int_{-\frac{\pi}{n}}^{\frac{\pi}{n}}d\phi\, e^{i\phi h}
  = \frac{n}{\pi h} \sin\left(\frac{\pi h}{n}\right)\label{eqn: volumn definition}
\end{align}
with $h=1,2,\hdots$ being integers and the normalization constant $V_{0} = \int_{\Omega} d\epsilon\, \epsilon^0 = \frac{2\pi}{n}$.

Extending this to non-abelian groups, e.g. $SU(N)$, $\Omega$ becomes a high-dimensional polytope in $SU(N)$ space. In~\cite{Flyvbjerg:1984dj}, the $V_r$ for $\bi$ and $\ds$ were computed up to second order by approximating these polytopes with hyperspheres to two significant figures. It is crucial to remove these approximations for our purpose because the uncertainty $\delta V_r\sim\mathcal{O}(1\%)$ is magnified in the coupling constants of the decimated action. These couplings are combinations of powers of $V_r$ with extreme cancellations making the fraction errors grow rapidly. Hence we avoid the hypersphere approximation and numerically compute all the $ V_r$ necessary for the 3rd order actions to $\mathcal{O}(0.1\%)$. (Results found in Table~\ref{tab:vrnum}.)

\begin{table*}
\caption{\label{tab:vrnum} The dimension, $d_r$, the character $\chi_r$, and $ V_r[G\rightarrow H]=d_r^{-1}\langle\Re \chi_r\rangle$ of character $r$ for the decimations $U(1)\rightarrow \zn$, $SU(2)\rightarrow \bi$, and $SU(3)\rightarrow \ds$. We have followed the normalizations in Table 14 of~\cite{Drouffe:1983fv}.}
\begin{center}
\begin{tabular}
{c | c c | c c c}
\hline\hline
$r$& $d_r$ & $\chi_r$ & $ V_r[U(1)\rightarrow \zn]$ & $ V_r[SU(2)\rightarrow \bi]$ & $ V_r[SU(3)\rightarrow \ds]$\\
\hline
$\{1\}$ & $N$ &$\Tr (U) $ &$\frac{n}{\pi} \sin\left(\frac{\pi}{n}\right)$ & 0.964748(2) &0.83414(6) \\
$\{2\}$ & $\frac{N(N+1)}{2} $ &$\frac{1}{2}\left(\tr^2(U)+\tr (U^2)\right) $ &$\frac{n}{2\pi} \sin\left(\frac{2\pi}{n}\right)$ & 0.90798(3) &0.62874(11) \\
$\{1,1\}$ & $\frac{N(N-1)}{2} $ &$\frac{1}{2}\left(\tr^2 (U)-\tr (U^2)\right) $ &--- & 1&0.83414(6) \\
$\{1,-1\}$ & $N^2-1 $ &$\left|\tr (U)\right|^2-1 $ &--- & 0.90798(3)&0.65971(10) \\
$\{3\}$  & $\frac{N(N+1)(N+2)}{6} $ &$\frac{1}{6}\left(\tr^3 (U)+2\tr(U^3)+3\tr(U^2)\tr(U)\right) $ &$\frac{n}{3\pi} \sin\left(\frac{3\pi}{n}\right)$ &0.83257(2) &0.42119(13) \\
$\{2,1\}$ &$\frac{N(N^2-1)}{3} $ &$\frac{1}{3}\left(\tr^3 (U)-\tr(U^3)\right) $ &--- & 0.964748(2)&0.65971(10) \\
$\{1,1,1\}$ & $\frac{N(N-1)(N-2)}{6} $ &$\frac{1}{6}\left(\tr^3 (U)+2\tr(U^3)-3\tr(U^2)\tr(U)\right) $ &--- &--- &1 \\
$\{2,-1\}$ & $\frac{N(N-1)(N+2)}{2} $ &$\frac{1}{2}\left(\tr^2(U)\tr (U^\dag)+\tr (U^2)\tr (U^\dag)\right)-\tr (U) $ &--- &0.83257(2) &0.46693(13) \\
$\{1,1,-1\}$ & $|\frac{N(N+1)(N-2)}{2}|$ & $\frac{1}{2}\left(\tr^2(U)\tr (U^\dag)-\tr (U^2)\tr (U^\dag)\right)-\tr (U) $ &$-\frac{n}{\pi} \sin\left(\frac{\pi}{n}\right)$ &--- &0.62874(11) \\
\hline\hline
\end{tabular}
\end{center}
\end{table*}

\section{Order-by-Order Decimation}
\label{sec:obo}

In this section, we summarize the derivation of the decimated action order-by-order.  Further details can be found in Appendix~\ref{apx:deriv}. The first order is relatively straight-forward, and only contains a single plaquette term.  Working from Eq.~(\ref{eq:first_order})
\begin{align}
  &\beta \mathscr{S}_1[u]= -\frac{\beta}{N}\langle \Re \tr{\left(u_1\epsilon_1u_2\epsilon_2(u_3\epsilon_3)^\dag(u_4\epsilon_4)^\dag\right)} \rangle \\
  &= -\frac{\beta}{N}\Re({u_1}_{ab}{u_2}_{cd}{u_3}_{ef}^\dag {u_4}_{gh}^\dag\langle {\epsilon_{1}}_{bc}\rangle\langle{\epsilon_{2}}_{de}\rangle\langle{\epsilon_{3}}_{fg}^\dag\rangle\langle{\epsilon_{4}}_{ha}^\dag \rangle).\notag
  \end{align}
  After applying Eq.~(\ref{eq:eij}), $\mathscr{S}_1[u]$ depends only on $u$:
  \begin{align}
  \label{eq: first-order-presentation}
  \beta \mathscr{S}_1[u]&= -V_{\{1\}}^4\frac{\beta}{N}\Re({u_1}_{ab}{u_2}_{cd}{u^\dag_3}_{ef}{u^\dag_4}_{gh})\delta_{bc}\delta_{de}\delta_{fg}\delta_{ha}\nonumber\\
&=-V_{\{1\}}^4\frac{\beta}{N}\Re\chi_{\{1\}}\upfun \equiv -\beta_{\{1\}}^{(1)} \frac{1}{N}\Re\chi_{\{1\}}\upfun \,,
\end{align}
where $\beta_r^{(n)}$ is the $n$-th order term in front of $\frac{1}{d_r}\Re\chi_r$. 

It is comforting that at $\mathcal{O}(\beta)$, no new terms are generated in $S[u]$. This allows for rescaling $\beta_{\{1\}}^{(1)}\rightarrow \beta$, recovering the procedure of directly replacing $U\rightarrow u$ in the Wilson action. Although this rescaling is permitted, $V_{\{1\}}<1$ contains content about the approximation $G\rightarrow H$. As the number of elements of $H$ increases, $\Omega$ shrinks and $V_{\{1\}}\rightarrow1$. This means $V_{\{1\}}$ quantifies how densely $H$ covers $G$ and thus the minimal fluctuation size. Since $\beta_{\{1\}}^{(1)}=V_{\{1\}}^4\beta$, decreases in $V_{\{1\}}$ signals the poorness of using Eq.~\eqref{eq: first-order-presentation} alone. This is discussed further in Sec.~\ref{sec:cont}.

\begin{table*}
\caption{\label{tab:betas} $\beta_r[G\rightarrow H]$ of character $r$ for a general group decimation. For completeness, we have included the 4 two-plaquette terms derived in~\cite{Flyvbjerg:1984ji,Flyvbjerg:1984dj} at second order labeled as $2r, 2i, 2t$ and $2u$.}
\begin{center}
\begin{tabular}
{c | c }
\hline\hline
$r$& $\beta_r$\\
\hline
$\{0\}$ &$\frac{1}{4N^2}[1-V_{\{1\}}^8]\beta^2$ \\
$\{1\}$ & $V^4_{\{1\}}\beta+\frac{1}{8N^2}V_{\{1\}}^4[4V_{\{1\}}^8-V_{\{1,1\} }^4-2V_{\{1,-1\}}^4-V_{\{2\}}^4]\beta^3 $ \\
$\{2\}$ & $ \frac{N+1}{8N}[V_{\{2\}}^4-V_{\{1\}}^8]\beta^2$ \\
$\{1,1\}$ & $\frac{N-1}{8N}[V_{\{1,1\}}^4-V_{\{1\}}^8]\beta^2 $\\
$\{1,-1\}$ &$\frac{N^2-1}{4N^2}[V_{\{1,-1\}}^4-V_{\{1\}}^8]\beta^2$\\
$\{3\}$ & $\frac{(N+1)(N+2)}{6N^2}[\frac{1}{24}V_{\{3\}}^4+\frac{1}{12}V_{\{1\}}^{12}-\frac{1}{8}V_{\{1\}}^{4}V_{\{2\}}^4]\beta^3 $\\
$\{2,1\}$ &$\frac{(N^2-1)}{6N^2}[\frac{1}{6}V_{\{2,1\}}^4+\frac{1}{3}V_{\{1\}}^{12}-\frac{1}{4}V_{\{1\}}^{4}V_{\{1,1\}}^4-\frac{1}{4}V_{\{1\}}^{4}V_{\{2\}}^4]\beta^3 $\\
$\{1,1,1\}$ &$\frac{(N-1)(N-2)}{6N^2}[\frac{1}{24}V_{\{1,1,1\}}^4+\frac{1}{12}V_{\{1\}}^{12}-\frac{1}{8}V_{\{1\}}^{4}V_{\{1,1\}}^4]\beta^3 $\\
$\{2,-1\}$ &$\frac{(N-1)(N+2)}{16N^2}[V_{\{2,-1\}}^4+2V_{\{1\}}^{12}-V_{\{1\}}^{4}(2 V_{\{1,-1\}}^4+V_{\{2\}}^4)]\beta^3 $\\
$\{1,1,-1\}$ &$|\frac{(N+1)(N-2)}{16N^2}|[V_{\{1,1,-1\}}^4+2V_{\{1\}}^{12}-V_{\{1\}}^{4}(2 V_{\{1,-1\}}^4+V_{\{1,1\}}^4)]\beta^3 $\\
\hline
{$\{2r\}$}&$\frac{1}{2}V_{\{1\}}^6[\frac{1}{4}V_{\{2\}}+\frac{1}{4}V_{\{1,1\}}+\frac{1}{2}V_{\{1,-1\}}-V_{\{1\}}^2]\beta^2$\\
 {$\{2i\}$}&$\frac{1}{2}V_{\{1\}}^6[\frac{1}{4}V_{\{2\}}+\frac{1}{4}V_{\{1,1\}}-\frac{1}{2}V_{\{1,-1\}}]\beta^2$\\
 $\{2t\}$&$\frac{1}{8N}V_{\{1\}}^6[V_{\{2\}}-V_{\{1,1\}}]\beta^2$\\
 $\{2u\}$&$\frac{1}{4N^2}V_{\{1\}}^6[1-V_{\{1,-1\}}]\beta^2$\\
\hline\hline
\end{tabular}
\end{center}
\end{table*}

We now proceed to calculate the second order decimated action while fixing a few typos in \cite{Flyvbjerg:1984dj} along the way. The second order decimated action $\mathscr{S}_2[u]=-\vev{\suep^2}+\vev{\suep}^2$ depends upon two plaquettes $U_{p}=U_1U_2U_3^\dag U_4^\dag$ and $U_{q}=U_5U_6U_7^\dag U_8^\dag$. 
A natural decomposition of $\mathscr{S}_2[u]$ can be made into three terms based on how the two plaquettes $p$ and $q$ are related: $p=q$ (one-plaquette contribution), $p\cap q= 1$-link (two-plaquette contribution), and $p\cap q=0$-links. To all orders, the $p\cap q=0$ contributions to the decimated action vanish.
%Absorbing an overall minus sign in Eq. (10) into the second order action, [Yao Comments: No minus sign is absorbed, $\mathscr S_2$ equals ...]
For the case of $p=q$, we conclude that it is:
\begin{align}
\label{eq:second-order-full}
  {\frac{1}{2!}}\beta^2\mathscr{S}_2[u]_{1p} =& -\beta_{\{0\}}^{(2)}-\beta_{\{2\}}^{(2)}\frac{2}{N(N+1)}\Re\chi_{\{2\}}\upfun\nonumber\\
  &-\beta_{\{1,1\}}^{(2)}\frac{2}{N(N-1)}\Re\chi_{\{1,1\}}\upfun \nonumber\\
  &-\beta_{\{1,-1\}}^{(2)}\frac{1}{N^2-1}\chi_{\{1,-1\}}\upfun \, ,
\end{align}
where the $\beta_r^{(2)}$ can be found in Table~\ref{tab:betas}.

Next, we calculate the case of $p\cap q= 1$-link for the second order decimation. Contracting the $\delta$'s in Eqs.~(\ref{eq:eeij}) and (\ref{eq:eedij}), where unlike Eq.~(\ref{eq:21p}), we only identify one link as the same between the two plaquettes.  This leads to the following expression,
\begin{align}
  &{\frac{1}{2!}}\beta^2\mathscr{S}_2[u]_{2p}\\
  &=-\beta_{\{2r\}}\frac{1}{N}\Re\left[\chi_{\{1\}}(u_p)\right]\frac1N\Re\left[\chi_{\{1\}}(u_q)\right]\notag\\
  &\quad-\beta_{\{2i\}}\frac{1}{N}\Im\left[\chi_{\{1\}}(u_p)\right]\frac{1}{N}\Im\left[\chi_{\{1\}}(u_q)\right]\notag\\
  &\quad-\beta_{\{2t\}}\frac{1}{N}\Re\left[\chi_{\{1\}}(u_{p*q^\dagger})\right]-\beta_{\{2u\}}\frac{1}{N}\Re\left[\chi_{\{1\}}(u_{p*q})\right]\,,\notag
\end{align}
where we have used the fact that all the $V_r$'s are real due to our choice of the integration region. The explicit expressions for the couplings are found in Table~\ref{tab:betas}.  Note that this expression is also applicable to $U(1)$.

\begin{figure}
  \centering
  \includegraphics[width=0.45\linewidth]{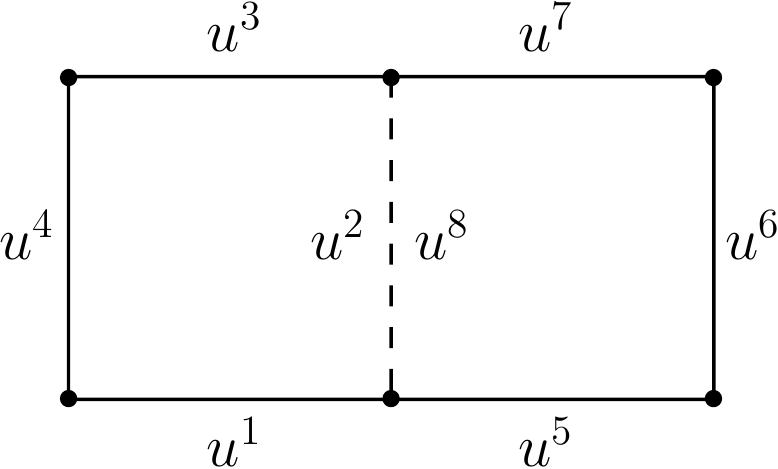}
  \includegraphics[width=0.48\linewidth]{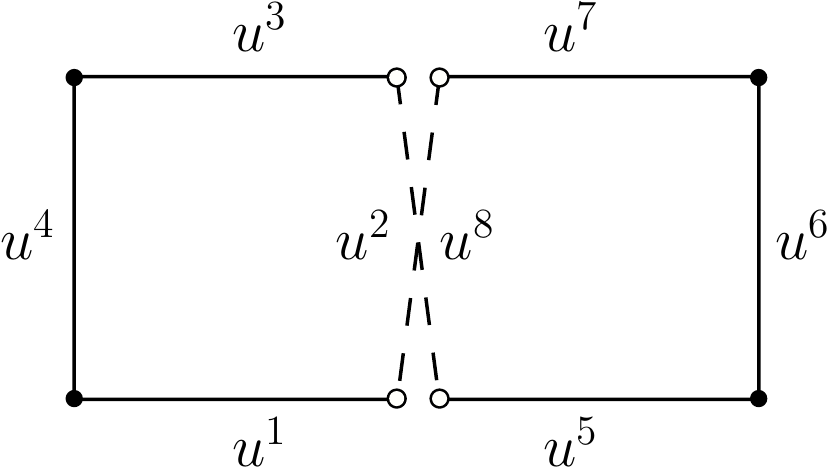}
  \caption{Example of two plaquettes $u_p$ and $u_q$ where $p\cap q=u_2=u_8$. The second order contributions depend on (top) $u_{p*q}=u_1u_5u_6u_7^\dagger u_3^\dagger u_4^\dagger$ and (bottom) $u_{p*q^\dag}=u_1u_2u_7u_6^\dagger u_5^\dagger u_2 u_3^\dagger u_4^\dagger$.}
  \label{fig:2plaq}
\end{figure}

We would now like to comment on how the two-plaquette -- and general multiplaquette -- terms contributes to the $S[u]$. It would be desirable if these terms could be neglected, because they require substantial quantum resources. By inspecting Table~\ref{tab:numbetas}, one observes that the two-plaquette $\beta_{r}$ are $\mathcal{O}(0.1)$ or smaller than the single-plaquette terms.  The largest coupling, $\beta_{2i}$, multiples a term $\Im \chi_1 \Im \chi_1\approx 0$.  Strong cancellations are expected from correlations between the remaining terms (shown in Fig.~(\ref{fig:2plaq})) as evident by the observation $\beta_{2t}\approx -\beta_{2u}$.

It is reasonable to expect these individual reasons to persist at higher orders, suggesting that at a fixed order all multi-plaquette terms can be neglected compared to their $1-$plaquette counterpart.  But can we argue that the multi-plaquette terms generated at order ${\cal O}(\beta^{n-1})$ are still negligible when the ${\cal O}(\beta^{n})$ contribution is introduced?  
To do this, we look at the continuum limit of each term being introduced.  In this way, we recognize that the two-plaquette terms are related to the L\"uscher-Weisz action~\cite{Luscher:1984xn}.
$k$-plaquette terms corresponds to applying $2k-2$ derivatives to $a^4\langle \Tr FF\rangle$ and are thus $\mathcal{O}(a^{2k+2})$. Here $F$ is the field strength tensor. Combining this with the observation that for a coupling $\beta_j$ generated at $\mathcal{O}(\beta^n)$ has the scaling $\beta_{j}\approx 10^{-n}\beta^n$, we estimate that
\begin{equation}
    \frac{\langle \mathscr{S}_{m}^{k-plaq}[u]\rangle}{\langle \mathscr{S}_{n}^{\rm 1 plaq}[u]\rangle}\approx \left(\frac{10}{\beta}\right)^{n-m}\frac{a^{2k+2}\langle  D^{2k-2}(\Tr FF)\rangle}{a^4\langle \Tr FF\rangle}\,,
\end{equation}
where $D$ is a covariant derivative projected onto the lattice directions.
The combination of higher powers of $a$ and the associated expectation values of higher-dimensional operators should be sufficient to suppress the mild $\beta$ enhancement for $n>m$, at least for $\mathcal{O}(\beta^3)$ $S[u]$.  For these reasons, we will neglect higher order multi-plaquette terms.

For the third-order terms of Eq.~\eqref{eq:connected2}, we, therefore, only focus on the case where three plaquettes are identical. Combining Eqs.~(\ref{eq:first-term-thrid-order}),~(\ref{eq:second-term-thrid-order}), and ~(\ref{eq:third-term-thrid-order}) we arrive at the third order contribution to the single-plaquette decimated action 
\begin{align}
\label{eq:third-order-full}
   \frac{\beta^3}{3!}\mathscr{S}_3[u]=&-\frac{\beta_{\{3\}}^{(3)}}{d_{\{3\}}}\Re\chi_{\{3\}}\upfun - \frac{\beta_{\{2,1\}}^{(3)}}{d_{\{2,1\}}}\Re\chi_{\{2,1\}}\upfun\\ & - \frac{\beta_{\{1,1,1\}}^{(3)}}{d_{\{1,1,1\}}}\Re\chi_{\{1,1,1\}}\upfun-\frac{\beta_{\{2,-1\}}^{(3)}}{d_{\{2,-1\}}}\Re\chi_{\{2,-1\}}\upfun\nonumber \\
   & -\frac{\beta_{\{1,1,-1\}}^{(3)}}{d_{\{1,1,-1\}}}\Re\chi_{\{1,1,-1\}}\upfun - \frac{\beta_{\{1\}}^{(3)}}{d_{\{1\}}}\Re\chi_{\{1\}}\upfun \,,\notag
\end{align}
where the overall factor of $1/3!$ has been absorbed into the definition of $\beta_r^{(3)}$. Note that, unlike the second order results  where only certain decimation programs generate renormalization for existing terms, the third order $S^3[u]$ introduces corrections to  $\Re\chi_1$ for all $G\rightarrow H$.  Additionally, a number of the specific group identities in Eqs.~(\ref{eq:u1idents})-(\ref{eq:su3idents}) also lead to renormalization.

Putting together Eqs.~(\ref{eq: first-order-presentation}),~(\ref{eq:second-order-full}), and ~(\ref{eq:third-order-full}), the single-plaquette decimated action of Eq.~\eqref{eq:z-effective} to $\mathcal{O}(\beta^3)$ for a general gauge group is, 
\begin{widetext}
\begin{align}
\label{eq:fullaction}
  S[u]=&\sum_p -\left(\beta_{\{1\}}^{(1)}+\beta_{\{1\}}^{(3)}\right) \frac{1}{N}\Re(\chi_{\{1\}}\upfun ) - \beta_{\{0\}}^{(2)}-\beta_{\{2\}}^{(2)}\frac{2}{N(N+1)}\Re\chi_{\{2\}}\upfun-\beta_{\{1,1\}}^{(2)}\frac{2}{N(N-1)}\Re\chi_{\{1,1\}}\upfun \nonumber\\
  &\qquad -\beta_{\{1,-1\}}^{(2)}\frac{1}{N^2-1}\chi_{\{1,-1\}}\upfun-\beta_{\{3\}}^{(3)} \frac{6}{N(N+1)(N+2)}\Re\chi_{\{3\}}\upfun -\beta_{\{2,1\}}^{(3)}\frac{3}{N(N^2-1)}\Re\chi_{\{2,1\}}\upfun \nonumber \\
  &\qquad  -\beta_{\{1,1,1\}}^{(3)}\frac{6}{N(N-1)(N-2)}\Re\chi_{\{1,1,1\}}\upfun -\beta_{\{2,-1\}}^{(3)}\frac{2}{N(N-1)(N+2)}\Re\chi_{\{2,-1\}}\upfun \nonumber \\
   &\qquad -\beta_{\{1,1,-1\}}^{(3)}\frac{2}{N(N+1)(N-2)}\Re\chi_{\{1,1,-1\}}\upfun \,,
\end{align}

\end{widetext}
where $\beta_r$ are in Table~\ref{tab:betas}. Note that this $S[u]$ is correct for any $G\rightarrow H$. Referring to Eqs.~(\ref{eq:u1idents}),~(\ref{eq:su2idents}), and~(\ref{eq:su3idents}), for a given $G$ simplifications occur. For $SU(3)$, with 
\begin{align}\label{eq:su3-beta-def}
\beta_r\equiv\sum_n \frac{1}{n!}\beta_r^{(n)}\, ,
\end{align}  this corresponds to:
\begin{align}
\label{eq:s3su3}
   S[u]=\sum_p&-\left(\beta_{\{1\}}+\beta_{\{1,1\}}\right)\frac{1}{3}\Re\chi_{\{1\}}\upfun  -\left(\beta_{\{0\}}+\beta_{\{1,1,1\}}\right)\nonumber\\&-\left(\beta_{\{2\}}+\beta_{\{1,1,-1\}}\right)\frac{1}{6}\Re\chi_{\{2\}}\upfun\nonumber\\&-\left(\beta_{\{1,-1\}}+\beta_{\{2,1\}}\right)\frac{1}{8}\chi_{\{1,-1\}}\upfun\nonumber \\
   &-\frac{\beta_{\{3\}}}{10}\Re\chi_{\{3\}}\upfun-\frac{\beta_{\{2,-1\}}}{15}\Re\chi_{\{2,-1\}}\upfun \,. 
\end{align}
For $U(1)$ and $SU(2)$, we refer the reader to Appendix~\ref{apx:groups}.

\begin{table*}[ht]
\caption{\label{tab:numbetas} Numerical values of $\beta_r[G\rightarrow H]$ of character $r$ for the decimations $U(1)\rightarrow \mathbb{Z}_{4}$, $SU(2)\rightarrow \bi$, and $SU(3)\rightarrow \ds$. For completeness, we have included the 4 two-plaquette terms derived in~\cite{Flyvbjerg:1984ji,Flyvbjerg:1984dj} at second order.}
\begin{center}
\begin{tabular}
{c | c c c}
\hline\hline
$r$& $\beta_r[U(1)\rightarrow \mathbb{Z}_{4}]$ & $\beta_r[SU(2)\rightarrow \bi]$ & $\beta_r[SU(3)\rightarrow \ds]$\\
\hline
$\{0\}$ & $0.142081\beta^2$ & $0.0155979(8)\beta^2$ & $0.021267(4)\beta^2+0.0008079(3)\beta^3$ \\
$\{1\}$ & $0.657022\beta+0.128321\beta^3$ &$0.866276(8)\beta-0.001350(8)\beta^3$ & $0.48411(13)\beta+0.020812(14)\beta^2-0.000550(4)\beta^3$\\
$\{2\}$ & $-0.066855\beta^2$ &$-0.02652(3)\beta^2$ & $-0.01301(3)\beta^2-0.000960(5)\beta^3$ \\
$\{1,1\}$ &--- &--- & --- \\
$\{1,-1\}$ &--- &--- &$-0.00999(4)\beta^2-0.001202(5)\beta^3$ \\
$\{3\}$ & $0.010483\beta^3$ &$0.001185(14)\beta^3$ & $0.000485(3)\beta^3$ \\
$\{2,1\}$ &--- &--- &--- \\
$\{1,1,1\}$ & --- & --- & --- \\
$\{2,-1\}$ & --- &--- & $0.001070(11)\beta^3$ \\
$\{1,1,-1\}$ & --- &--- &--- \\
\hline
 $\{2r\}$ & $-0.173459\beta^2$  & $0.000102(7)\beta^2$ & $-0.000034(18)\beta^2$ \\
 $\{2i\}$& $0.04238\beta^2$  & $0.009295(7)\beta^2$ & $0.006040(10)\beta^2$ \\
 $\{2t\}$& $0.04238\beta^2$  & $-0.0046477(15)\beta^2$  & $-0.002883(19)\beta^2$\\
 $\{2u\}$& $0.13314\beta^2$  & $0.0046477(15)\beta^2$ & $0.003184(15)\beta^2$\\
\hline\hline
\end{tabular}
\end{center}
\end{table*}

\section{Results for $\ds$}
\label{sec:num}
\begin{figure}
  \centering
  \includegraphics[width=0.95\linewidth]{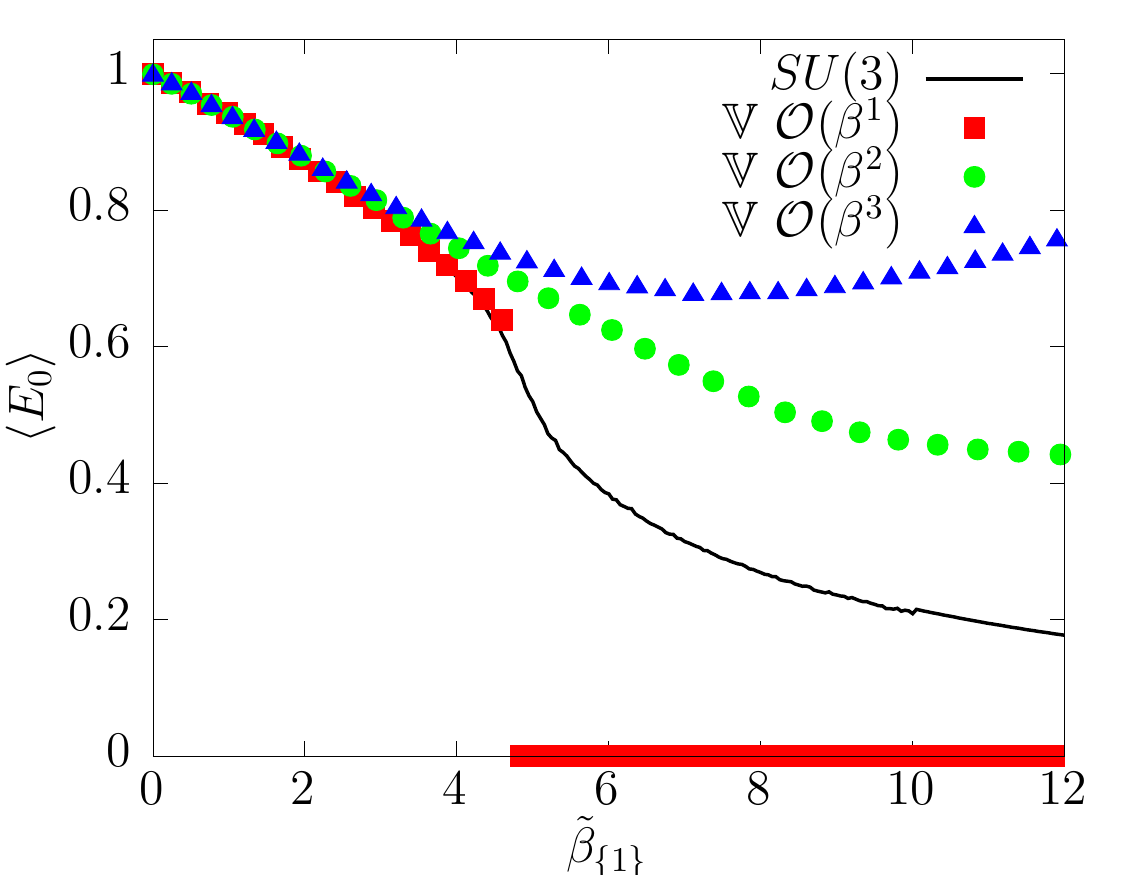}
  \caption{Average energy per plaquette, $\langle E_0\rangle = 1-\Re\langle\Tr U_p\rangle /3$, vs $\tilde{\beta}_1$ on $4^4$ lattice for $\ds$ action with corrections of: (\protect\redsq) $\mathcal{O}(\beta^1)$,  (\protect\gcir) $\mathcal{O}(\beta^2)$, and (\protect\bltri) $\mathcal{O}(\beta^3)$. The black line is the $SU(3)$ result.}
  \label{fig:plaqv}
\end{figure}

As a demonstration, we simulated Eq.~(\ref{eq:s3su3}) to each order in $\beta$ for $SU(3)\rightarrow\ds$.  For these computations 10$^2$ configurations separated by 10$^3$ sweeps were collected on a $4^4$ lattice and plotted in Fig.~\ref{fig:plaqv}. In the figure, we compare the average energy per plaquette $\langle E_0\rangle$ versus the coupling, ${\beta}_{\{1\}}$ as defined in Eq.~\eqref{eq:su3-beta-def}, which multiplies the Wilson term $\Re\chi_1$. For $SU(3)$, this corresponds to $\beta$, and $\beta_{\{1\}}+\beta_{\{1,1\}}$ for $\ds$. 

Naively, $\langle E_0\rangle$ is monotonic in $a$. Including $\mathcal{O}(\beta^2)$ terms, we observe a clear reduction in $\langle E_0\rangle$ and thus an improvement over the Wilson action for small $a$.  This suggests promise in this systematic approach. As will be discussed in Sec.~\ref{sec:cont}, it also supports the effectiveness of the previously studied ad-hoc actions. Instead of freezing out, the theory approaches a non-zero value of $\langle E_0\rangle$.

At $\mathcal{O}(\beta^3)$, $\langle E_0\rangle$ displays non-monotonic behavior due to the negative coefficient of $\chi_1$ at $\beta^3$.  This suggests the higher order terms (4th order and beyond) required to match Eq.~\eqref{eq:z-natural-expansion} and \eqref{eq:z-effective} are dominating the action.

Taken together, the $\mathcal{O}(\beta^n)$ results suggests that the decimation procedure, formulated as a strong coupling expansion, converges to the continuous theory slowly. Sufficiently large order calculations would suppress the higher order contributions in the form $\beta^n/n!$ for a reasonable range of $\beta$. While an $\mathcal{O}(\beta^4)$ calculation would undoubtedly be insightful to understanding this convergence, other approaches such as introducing ``counter-terms" to absorb some higher-order contributions, or instead using a character expansion may prove fruitful. It also may prove useful to take the expansion in Eq.~\eqref{eq:z-effective} as an effective theory where each character is subject to field redefinitions. While these possibilities are interesting to investigate, they are certainly beyond the scope of this work. We therefore leave them for future studies.

\section{Finite group effects}
\label{sec:cont}
With Eq.~(\ref{eq:fullaction}), it is possible for us to investigate systematically the effect of replacing the continuous group by its finite subgroup. 
In order to proceed, it is useful to introduce a new parameter which approximately represents the field fluctuations. To do this, consider the representation of a continuous group lattice gauge link in terms of the corresponding generators $\lambda_a$ in the adjoint representation, $U=e^{i\lambda_a A_a}$, where a summation over color indices $a$ is implied. In this form, we see that the gauge fields correspond to amplitudes in each of the generators. 
For $\epsilon\in \Omega$, inserting its small parameter expansion $\epsilon\approx \mathbb{1}+i \lambda_a A_a-\frac{1}{2}(\lambda_a A_a)^2+\hdots$ into Eq.~\eqref{eq:v_r_definition} gives 
\begin{equation}
  V_r \approx 1-\int_\Omega DA\left(c_r^{(2)} \sum_a A_a^2+\hdots\right)\, ,
\end{equation}
where $DA$ is a measure over all $A_a$ which respects gauge symmetry and $c_r^{(n)}$ are representation and group-dependent constants. From this, we see that as the subgroup $H$ incorporates more elements, the size of $\Omega$ approaches 0 and $ V_r\rightarrow 1$ from below. This means that for finite $\Omega$ the domain size of $A_a$ that grives rise to $\Omega$ is an indicator for deviations from $G$ of $H$. Flyvbjerg defines a parameter $R$ as the radius of a hypersphere with equal volume to $\Omega$ to get a handle on the domain of $A_a$. This allows him to approximate $V_{r}$ analytically~\cite{Flyvbjerg:1984dj,Flyvbjerg:1984ji}. Here, we can use this idea to roughly understand the scaling of $V_r$.

For $U(1)\rightarrow \zn$, the hypersphere is exactly $\Omega$ and $R$ cleanly defines $\epsilon\leq R=\pi/n$. Beyond $U(1)$, the connection between $\Omega$ and a single value of $R$ is complicated because the $\Omega$ of $H$ form polytopes in the hypervolume of their continuous partner (see Fig.~\ref{fig:sphere} for a clear demonstration). 
In this case, while one could take $\Omega$ to be contained by a hypersphere  centered at $\mathbb{1}$ whose boundary incorporates elements of the nearest neighbors of ${\mathbb 1}$ in $H$, making some element of the hypersphere not included in $\Omega$. 
On the other hand, there exists a largest hypersphere centered at $\mathbb{1}$ that \textit{only} contains elements in $\Omega$. 
In this way, we define an upper and lower bound for $R$. Note, this is different from~\cite{Flyvbjerg:1984dj,Flyvbjerg:1984ji} where the polytopes of $H$ were always approximated by hyperspheres with definite radii. 
For SU(2) with $\bi$, we find $0.09\leq R^2 \leq0.15$ which can be compared to $R^2_{sphere}=0.12$ of ~\cite{Flyvbjerg:1984dj,Flyvbjerg:1984ji}. In the case of SU(3) with $\ds$, $0.42\leq R^2 \leq 0.93$ compared to $R^2_{sphere}=0.62$. 

While superficially the cumulant expansion has appeared as a strong-coupling expansion in $\beta$, the actual behavior is controlled by both $\beta$ and $R$ with $R$ controlling $V_r$. As pointed out in~\cite{Flyvbjerg:1984dj,Flyvbjerg:1984ji}, the leading order behavior for small $R$ for a given power of $\beta^\alpha$ ($\alpha>0$) is actually $\mathcal{O}([\beta R^{2}]^\alpha R^{-2})$. Therefore one would predict that the relative smallness of $R^2$ for $\bi$ compare to $\ds$ signals that $\beta_f$ should be larger for $\bi$ which is indeed the case.

For subgroups of $SU(3)$, this scaling behavior becomes unsatisfactory because $R^2\sim 1$. It is possible to study this breakdown in $U(1)\rightarrow\zn$ where the systematic effect of decimation can be studied in detail both because errors can be made arbitrarily small for large $n$ and because  $ V_r$ and $\beta_r$ are known analytically. In terms of $R$, one can expand the $\beta_r$ for the $U(1)$ action of Eq.~\eqref{eq: u1_action} to find:
\begin{align}
  \beta_{\{0\}}&\approx\bigg(\frac{R^2}{3}-\frac{19R^4}{90}+\hdots\bigg)\beta^2, 
\end{align}
\begin{align}
  \beta_{\{1\}}+\beta_{\{1,1,-1\}}&\approx\left(1-\frac{2R^2}{3}+\frac{R^4}{5}+\hdots\right)\beta \nonumber\\&\quad+ \left(-\frac{17R^4}{90}+\frac{311R^6}{945}+\hdots\right)\beta^3,
  \end{align}
\begin{align}
  \beta_{\{2\}}&\approx\left(-\frac{R^2}{3}+\frac{53R^4}{90}+\hdots\right)\beta^2,  \end{align}
\begin{align}
  \beta_{\{3\}}&\approx \bigg(\frac{17R^4}{90}-\frac{1609R^6}{2835}+\frac{46303R^8}{56700}\nonumber\\&\quad\quad-\frac{77603R^{10}}{103950}+\hdots\bigg)\beta^3.
\end{align}
The first thing to note is that the $\mathcal{O}([\beta R^{2}]^\alpha R^{-2})$ scaling found in~\cite{Flyvbjerg:1984dj,Flyvbjerg:1984ji} continues to the third order. One might be tempted to use this leading behavior to estimate the $\beta_f$ or the radius of convergence of this series, but this would be incorrect. 
Instead, it behooves one to note that for both 2nd and 3rd order contributions, the subleading terms $[R^{2}]^k R^{-2}$ with $k>\alpha$ initially grow until a $1/k!$ factor dominates over all the other factors. 

But what is the origin of this behavior? 
For simplicity, we can understand this behavior by considering the expansion of $V_{j}^{4m}$ which form $\beta_r$. The specific combination of $V_{j}^{4m}$ dictated by the cumulant expansion ensures that orders lower than $\mathcal{O}([\beta R^{2}]^\alpha R^{-2})$ cancel in $\beta_r$. The $j$ representation contributes to $\beta_{\{r_1,\cdots, r_k\}}$ in the form of $V_{\{j_1^1,\cdots, j_l^1\}}^{4m_1}\cdots V_{\{j_1^k,\cdots,j_n^k\}}^{4m_k}$ under the constraint $|r|=m_i\widetilde j_i$ where $|r|=|r_1|+\cdots+|r_k|$ and $\widetilde j_i=|j_1^i|+\cdots+|j_l^i|$.
One might worry that studying the expansion of $V_{j}^{4m}$ isn't representative, but one can verify that the scaling behavior observed below persists in $\beta_r$, although the numerical factors become cumbersome. For $V_{j}^{4m}\equiv V_{\{j_1,\cdots,j_l\}}^{4m}$, $\widetilde j=|j_1|+\cdots |j_l|$, we have
\begin{equation}
V_{j}^{4m}\approx1-\frac{2}{3}m(\widetilde jR)^2+\frac{1}{45}(10m^2-m)(\widetilde jR)^4+\mathcal{O}(m^3[\widetilde jR]^6)
\end{equation}
from which, we see that the coefficients of the $[R^{2}]^k R^{-2}$ contributions to $\beta_r$ are accompanied by a factor $\propto\frac{1}{k!}m^{k-1} \widetilde j^{2k-2}$. While the factorials ensure the series converges, $\beta_r$ for higher representations $r$ have larger $m$, $\widetilde j$, or both leading to higher order terms in the expansion being large for moderate $R$. This helps explaining why ${\mathbb{Z}}_4$ with the Wilson action fails to replicate $U(1)$ substantially above $\beta=1$ -- while the naive scaling would suggest $R\lesssim\sqrt{\beta}^{\frac{k}{1-k}}$ would be enough to suppress higher representations, in reality a stronger bound of $\max\{\frac{1}{k!}m^{k-1} (\widetilde jR)^{2k-2}\}_{1\leq \widetilde j\leq |r|}\lesssim1$ for $\forall \beta_r$ is required for all subleading terms to be small.  Considering the range of $m$ with fixed $|r|,\widetilde j$, the bound is strictest when $|r|=\widetilde j$ yielding $R\lesssim1/|r|^{3/2}$ in order for the lowest order contribution to dominate such that $R\lesssim\sqrt{\beta}^{\frac{\alpha}{1-\alpha}}$ provides a reasonable estimate for the range of $\beta$ where the decimated action provides a reasonable approximation for its continuous partner. While these conditions are satisfied for $\bi$, they are violated for $\ds$ in which case the dominant term in the $R$ expansion isn't clear.

\begin{figure}
 \includegraphics[width=0.48\linewidth]{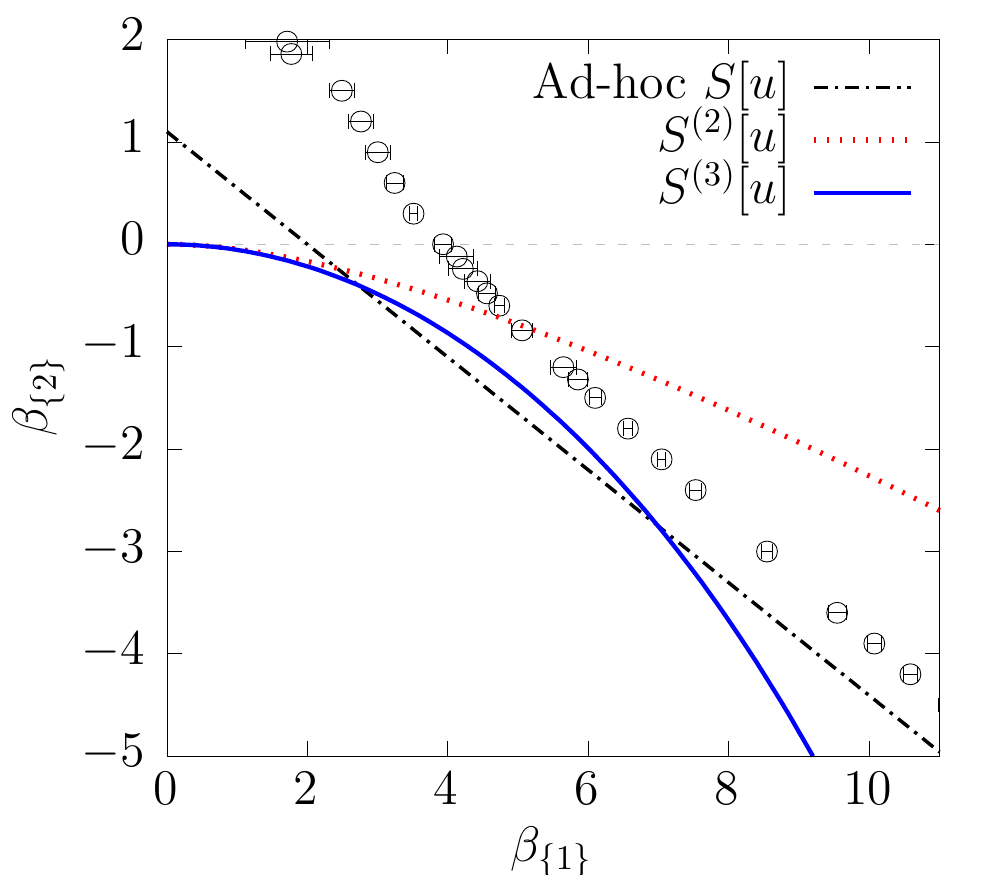}
 \includegraphics[width=0.48\linewidth]{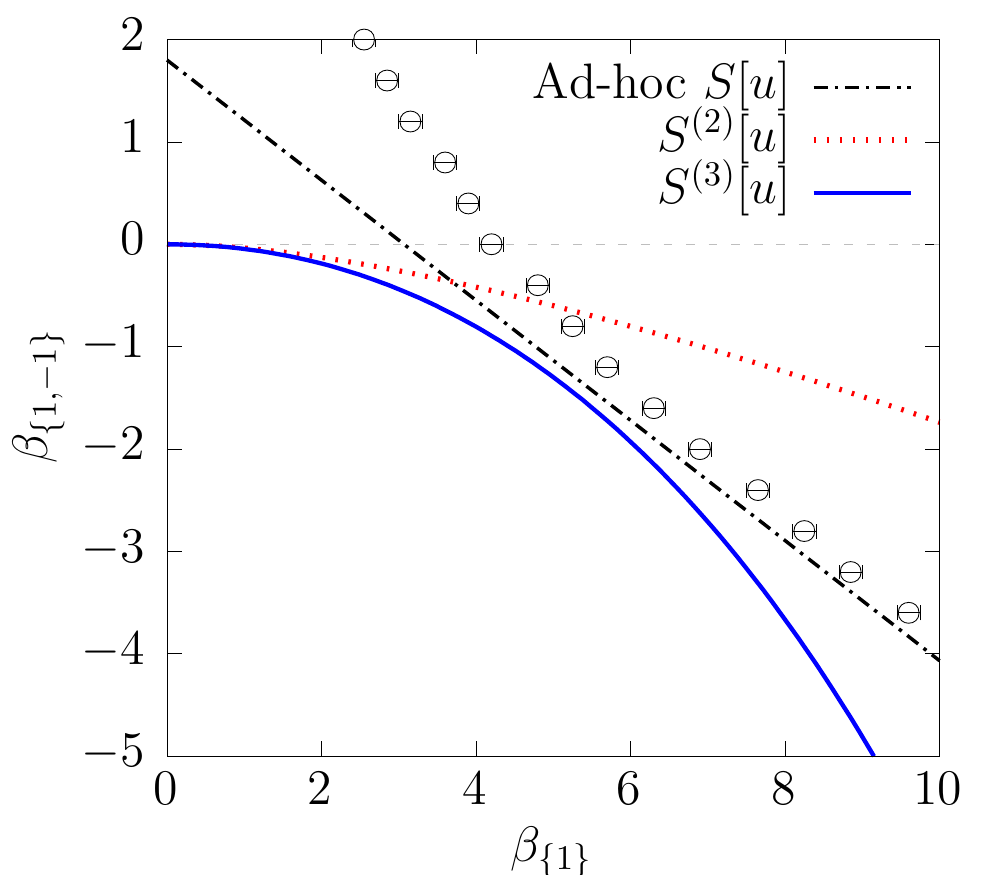}
\caption{\label{fig:effb}$\beta_{\{r\}}$ trajectories of $S^n[u]$ to the ad-hoc actions with additional terms (left) $\Re\chi_{\{2\}}$ of \cite{Alexandru:2019nsa} and (right) $\Re\chi_{\{1,-1\}}$ of~\cite{lammtocome}. The open circles indicate the boundary between the frozen and unfrozen phases obtained on $2^4$ lattices.}
\end{figure}

Another feature observed in the $R$ expansion of the $\zn$ group is that because $V_r\propto \sin rR$, the sign of the $\mathcal{O}([rR]^k)$ terms oscillate, and therefore the sign of $\beta_r^{(n)}$ can depend sensitively on $R$. Since $\Re\chi_r(\mathbb{1})>\Re\chi_r(\mathcal{N})$, where ${\cal N}$ is the nearest neighbors of ${\mathbb 1}$ in $H$ (see Fig.~\ref{fig:sphere}), the overall sign of $\beta_r^{(n)}$ determines whether or not the $r$-th term in the action enters the frozen phase in the limit of $\beta\rightarrow \infty$. 
This behavior is observed in Fig.~\ref{fig:effb} where $\beta_{\{1\}}^{(1,2)}>0$ but $\beta_{\{1\}}^{(3)}<0$.

From the behavior observed in $U(1)$, we can improve the quantitative understanding of how well $H$ can approximate $G$, even when $\beta_r$ are not known analytically. Clearly, $V_r\rightarrow1$ indicates that the $R\rightarrow0$, and in that limit the two actions would agree. Therefore, the difference between the two actions $S_G-S_H\approx\beta\chi_{\{1\}}(U)-\beta_{\{1\}}\chi_{\{1\}}(u)\approx (1-V_{\{1\}}^4)\beta\chi_{\{1\}}(u)$ serves as an indicator of $\beta_f$.

\begin{table}
\caption{\label{tab:groupcomp} Parameters of a discrete subgroups necessary to study the behavior of $\beta_f$.}
\begin{center}
\begin{tabular}
{c c| c c c c}
\hline\hline
$G$ & $H$& $\Delta S$ & $\mathcal{N}$ & $C$ & $V_{\{1\}}$ \\
\hline
$U(1)$&$\mathbb{Z}_{2}$&2&1&2&0.6366\\
&$\mathbb{Z}_{4}$&1&2&4&0.9003\\
&$\mathbb{Z}_{10}$&$\frac{3-\sqrt{5}}{4}$&2&10&0.9836\\
\hline
$SU(2)$&$\mathbb{BT}$&$\frac{1}{2}$&8&6&0.8939\\
&$\mathbb{BO}$&$\frac{1-\sqrt{2}}{2}$&6&8&0.9309\\
&$\bi$&$\frac{3-\sqrt{5}}{4}$&12&10&0.9648\\
\hline
$SU(3)$&$S(108)$&$\frac{2}{3}$&18&4&0.7138\\
&$S(216)$&$\frac{2}{3}$&54&4&0.7557\\
&$S(648)$&$1-\frac{1}{3}\left(\cos \frac{\pi}{9}+\cos \frac{2\pi}{9}\right)$&24&9&0.7855\\
&$\ds$&$\frac{5-\sqrt{5}}{6}$&72&5&0.8342\\
\hline\hline
\end{tabular}
\end{center}
\end{table}
\begin{figure*}
  \centering
  \includegraphics[width=0.27\linewidth]{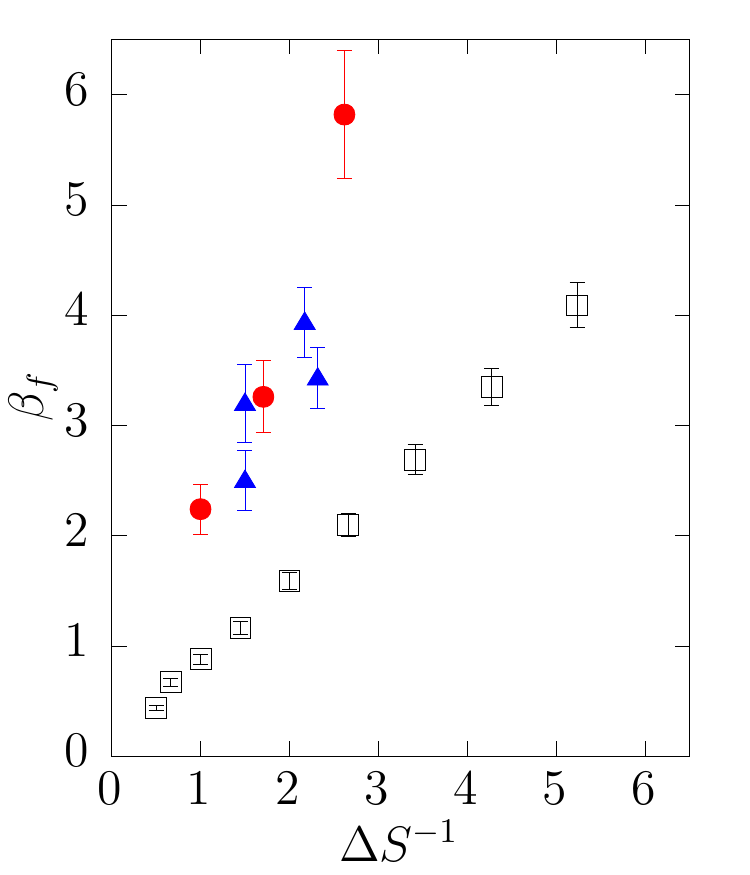}
  \includegraphics[width=0.27\linewidth]{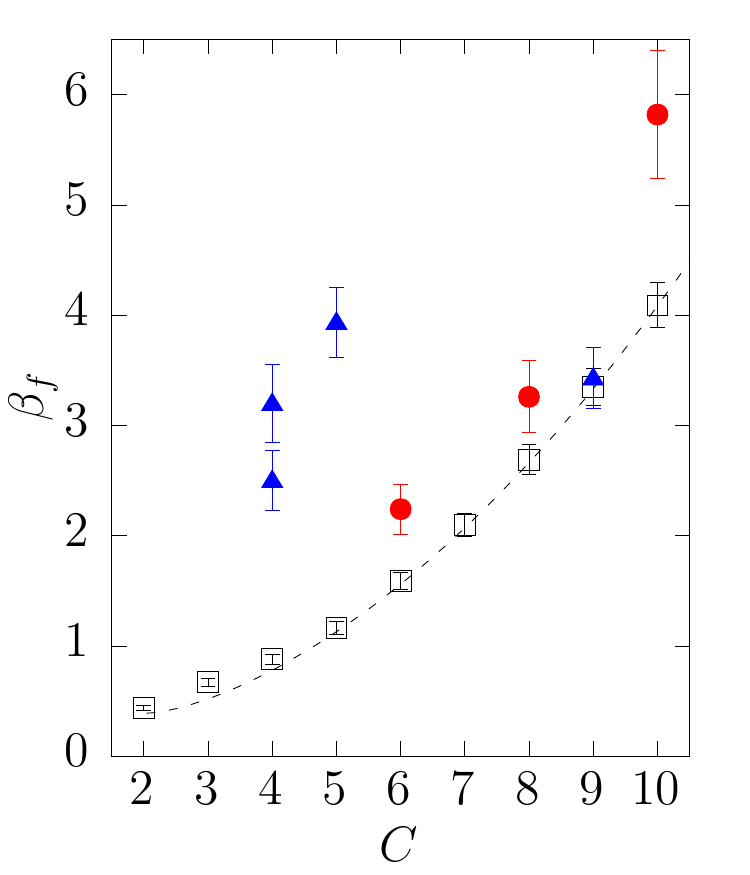}
  \includegraphics[width=0.27\linewidth]{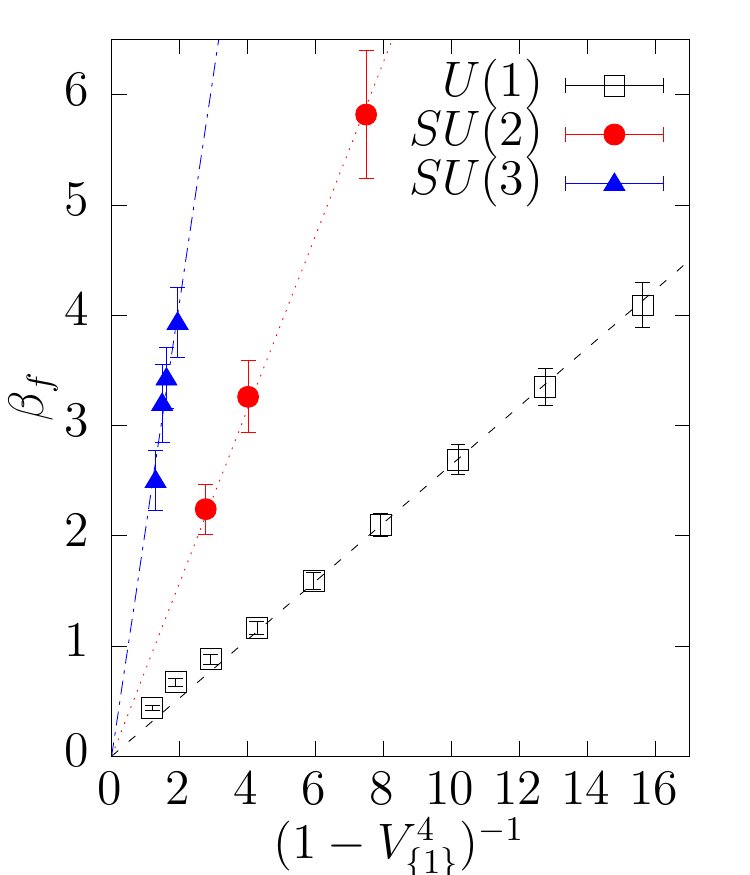}
  \caption{$\beta_f$ as a function of (left) $\Delta S^{-1}$, (center) the cycle $C$ of $\mathcal{N}$, (right) $(1-V_{\{1\}}^4)^{-1}$. Note that for the subgroups of $U(1)$ and $SU(2)$, monotonic behavior is observed for all three variables, but only for $(1-V_{\{1\}}^4)^{-1}$ are the subgroups of $SU(3)$ monotonic.}
  \label{fig:bcv1}
\end{figure*}
This proxy can be compared to others in the literature. The simplest estimate is $\beta_f^{-1}\propto\Delta S=\Re\tr(\mathbb{1})-\Re\tr(\mathcal{N})$~\cite{Petcher:1980cq}. While this estimate finds monotonic behavior for discrete groups of $U(1)$ and $SU(2)$, different $\mathcal{O}(1)$ factors are needed. It also fails completely for $SU(3)$, as seen in the left panel of Fig.~\ref{fig:bcv1}. 

Observing the differing $\mathcal{O}(1)$ factors, \cite{Petcher:1980cq} suggested a different estimate. For discrete Non-abelian subgroups near $\beta_f$, $S[u]$ is dominated by contributions from $u_p=\mathcal{N}$. 
From duality arguments
, the action near $\beta_f$ could be approximately rewritten as a $\mathbb{Z}_C$ action where $C$ is the minimal cycle such that $u^C=\mathbb{1}$ for all $u\subset \mathcal{N}$. Since $C=n$ for $\zn$, these arguments predict a single curve $\beta_f\approx0.78/(1-\cos(2\pi/C))$ directly from the study of $\beta_f$ in $\zn$ for all discrete subgroups. The discrepancy between $SU(2)$ and $U(1)$ was reduced from $\sim300\%$ to $\sim50\%$. The authors of \cite{Petcher:1980cq} warned that this approximation could be poor for $SU(3)$ albeit without numerical evidence. Since then $\beta_f$ for the subgroups of $SU(3)$ have been found and as anticipated, this estimator proves to be poor as presented in the center of Fig.~\ref{fig:bcv1}.
\begin{figure}
  \centering
  \includegraphics[width=0.9\linewidth]{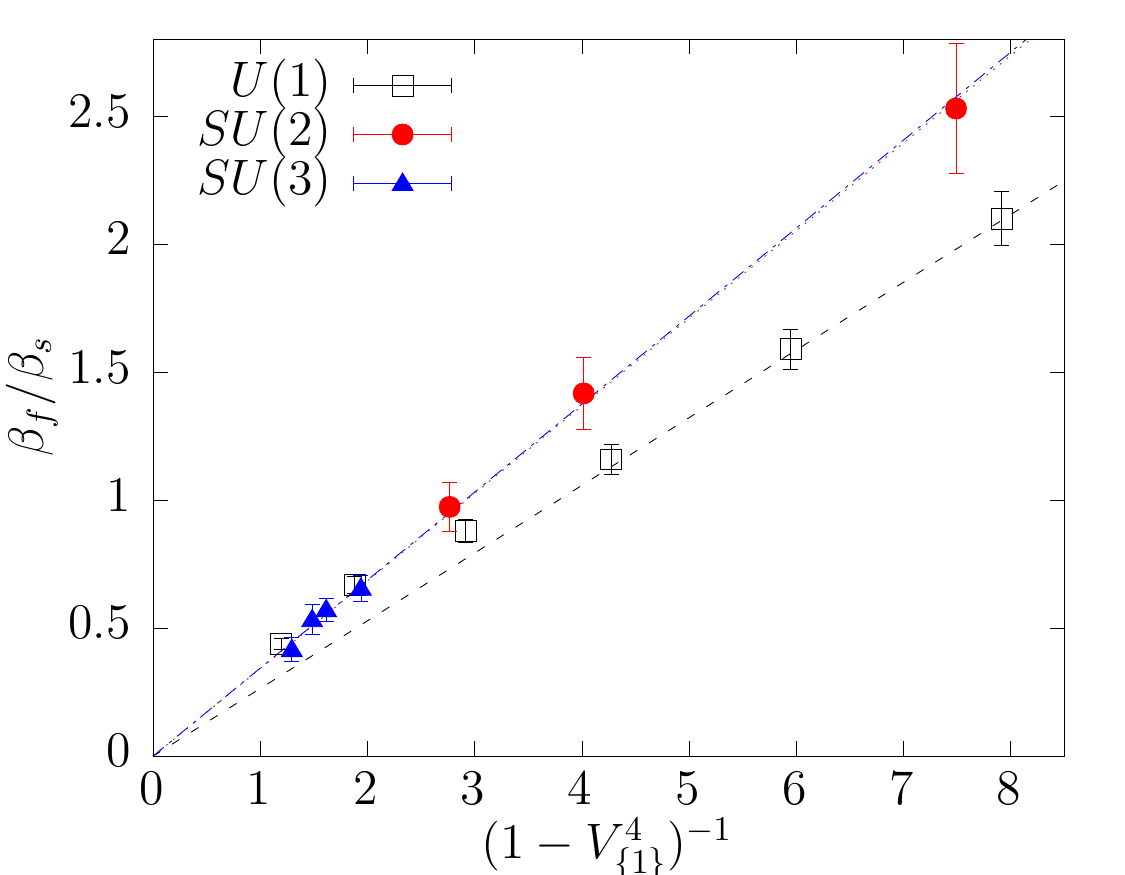}
  \caption{$\beta_f/\beta_s$ as a function of $(1-V_{\{1\}}^4)^{-1}$.}
  \label{fig:bcv1_scale}
\end{figure}
In the plot on the right of Fig.~\ref{fig:bcv1}, $\beta_f$ is plotted as a function of $(1-V_{\{1\}}^4)^{-1}$. We find that monotonic, linear behavior is observed within the uncertainties for each continuous group. Best fit lines have been included for each group to guide the eye. This suggests that our estimator captures some of the non-perturbative behavior near the freezing transition better than $\Delta S^{-1}$ or $C$. Physics of the different groups differ, as signaled by their different scaling regimes. If we divide $\beta_f$ by a rough estimate of $\beta_s=[1,2.2,6]$ for $U(1),SU(2),SU(3)$ respectively, we might expect to further remove some of this group dependence. Doing so in Fig.~\ref{fig:bcv1_scale}, we find that $SU(2)$ and $SU(3)$ collapse onto a single line and $U(1)$ within $~25\%$.

Using our higher order results, one can then gain insight into the effectiveness of the ad-hoc actions of $\ds$. Each of these actions corresponds to terms that are generated at 2nd order in the decimated action. The first ad-hoc action used in \cite{Alexandru:2019nsa} can be rewritten as
\begin{align}
 S[u]&=-\sum_p \left(\frac{\beta_0}{3}\Re\Tr (u_p) +\beta_1\Re\Tr(u_p^2)\right) \,,\nonumber\\
 &=-\sum_p \left(\left(\beta_0-3\beta_1\right)\frac{1}{3}\Re\chi_{\{1\}} +\left(6\beta_1\right)\frac{1}{6}\Re\chi_{\{2\}}\right) \,,\nonumber\\
 \end{align}
where we have used $\beta_1=a\beta_0+b$ with $a=-0.1267$ and $b=0.253$. For an unpublished action of 
\begin{equation}
 S[u]=-\sum_p\left(\frac{\tilde{\beta}_{\{1\}}}{3}\Re\chi_{\{1\}} +\frac{\tilde{\beta}_{\{1,1\}}}{8}\Re\chi_{\{1,-1\}}\right)
\end{equation}
where $\tilde{\beta}_{\{1,1\}}=a\tilde{\beta}_{\{1\}}+b$ with $a=-0.587$ and $b=1.80$. The trajectory parameters were chosen to be parallel to the freezing point at large $\beta_0$ by eye. From Fig.~\ref{fig:effb}, we see that in both ad-hoc actions, reasonably agreement is found for intermediate $\beta$ for the 3rd order action. Here $\beta$ is the coefficient in front of $\Re\chi_{\{1\}}$ for the ad-hoc actions. The ad-hoc trajectories are known to poorly reflect $G$ at low $\beta$, because they lack curvature to fix the known requirements at $\beta=0$. At large $\beta$, we expect higher order terms in the cumulant expansion to become relevant and thus disagreement is expected to occur. This surprising agreement in the intermediate region of $\beta$ suggests that actions formed by neglecting terms in the cumulant expansion are optimized in their character basis by setting the couplings to results given by the resuming higher order contributions in cumulant expansion with fluctuations $G/H$ integrated out.

\section{Conclusion}
\label{sec:concl}
In this work, we used the cumulant expansion to develop a systematic method for studying and improving lattice actions that replace continuous gauge groups by their discrete subgroups. This is a step in the ongoing trek toward developing accurate and efficient digitization on quantum computers. These decimated actions, through the factor $V_{\{1\}}$, have superior predictive power for finding the freezing transition compared to prior estimators.

We further computed the third-order, single-plaquette contribution for the general group. These higher-order terms are necessary for systematizing the decimation procedure of $SU(3)\rightarrow \ds$ where it has been observed that the inclusion of terms generated in the second-order cumulant expansion with ad-hoc couplings improve the approximation of $SU(3)$. The most immediate work in these directions would be to compute more Euclidean observables (i.e. Wilson flow parameter and pseudocritical temperature) from the full decimated action of Eq.~\eqref{eq:fullaction} and compare them to~\cite{Alexandru:2019nsa,lammtocome}. Given the large corrections from second to third order for $\ds$, additional work should be devoted to computing the fourth-order contributions.

In order to move beyond pure gauge theory, it will be necessary to consider quark fields. While the computational resources increase substantially for dynamical quarks, an advantage of the discrete subgroup approximations is that many standard lattice field theory techniques such as \emph{fermionic determinants} and \emph{pseudofermions} can be applied.  This was demonstrated in early works on dynamical fermions where $\bi$ replaced $SU(2)$~\cite{Weingarten:1980hx,Weingarten:1981jy}.

Another important step in studying the feasibility of this procedure is to explicitly construct the quantum registers and primitive gates \`a la \cite{Lamm:2019bik} where smaller discrete groups were investigated. Together with classical lattice results, this would allows for resource counts.

\begin{acknowledgments}
The authors would like to thank Scott Lawrence, Jesse Stryker, Justin Thaler, and Yukari Yamauchi for helpful comments on this work. Y.J. is grateful for the support of DFG, grants BR 2021/7-2 and SFB TRR 257. H.L. is supported by a Department of Energy QuantiSED grant. Fermilab is operated by Fermi Research Alliance, LLC under contract number DE-AC02-07CH11359 with the United States Department of Energy. S.Z. is supported by the National Science Foundation CAREER award (grant CCF-1845125).
\end{acknowledgments}

%%%%%%%%%%%%%%%%%%%%%%%%%%%%%%%%%%%
%%%%%%%%%%%%%%%%%%%%%%%%%%%%%%%%%%%

\bibliographystyle{apsrev4-1}
\bibliography{wise}

\appendix
\begin{widetext}
\section{Creutz Identities}
\label{apx:eident}
A useful identity was derived in~\cite{Creutz:1978ub} for $SU(N)$ and $U(N)$ groups such that for any integer $n\leq N$

\begin{align}
\vev{\epsilon_{i_1j_1}\cdots \epsilon_{i_nj_n}}&=c_1\,\varepsilon_{i_1A^1_1\ldots A^1_{N-1}}\varepsilon_{j_1A^1_1\ldots A^1_{N-1}}\times\cdots\times\varepsilon_{i_nA^n_1\ldots A^n_{N-1}}\varepsilon_{j_nA^n_1\ldots A^n_{N-1}}\notag\\
&+c_2\,\varepsilon_{i_1i_2A^1_1\ldots A^1_{N-2}}\varepsilon_{j_1j_2A^1_1\ldots A^1_{N-2}}\times\cdots\times\varepsilon_{i_n A^{n-1}_1\ldots A^{n-1}_{N-1}}\varepsilon_{j_n A^{n-1}_1\ldots A^{n-1}_{N-1}}+\cdots\notag\\
&+c_{B_n}\,\varepsilon_{i_1i_2\ldots i_n A^1_1\ldots A^1_{N-n}}\varepsilon_{j_1j_2\ldots j_n A^1_1\ldots A^1_{N-1}} \,,\label{eq:generalUU}
\end{align}

where $\varepsilon$ is Levi-Civita symbol, $A_i^j$ are the contracted dummy indices, and $B_n$ is the Bell number accounting for the number of ways that one can put the open indices $i_k\,,j_l$ on $\varepsilon$ such that no $i_k$ and $j_l$ appear in the same $\varepsilon$. In~\cite{Creutz:1978ub}, Eq.~\eqref{eq:generalUU} was derived for integrating over the entire group $G$. 
Hence in our case, we need to determine the constants $c_i$'s when integrating only over $\Omega$ for $\vev{\epsilon_{ij}}, \vev{\epsilon_{ij}\epsilon_{k\ell}}$, $\vev{\epsilon_{ij}\epsilon_{k\ell}^\dag}$, $\vev{\epsilon_{ij}\epsilon_{k\ell}\epsilon_{mn}}$, $\vev{\epsilon_{ij}\epsilon_{k\ell}\epsilon_{mn}^\dag}$, with $i,j,k,l,m,n\in [N]$.
This is done by contracting the tensor structure on each side of Eq.~\eqref{eq:generalUU} with products of Kronecker delta's and solving the resulting linear equations. 

At first order, only one integral is needed:
\begin{align}
\label{eq:eij}
\vev{\epsilon_{ij}}=&V_{\{1\}}\,\delta_{ij}\,.
\end{align}
At second order, there are two relations
\begin{align}
\label{eq:eeij}
\vev{\epsilon_{ij}\epsilon_{kl}}=&\frac{1}{2}\left(V_{\{2\}}+V_{\{1,1\}}\right)\delta_{ij}\delta_{kl}+\frac{1}{2}\left(V_{\{2\}}-V_{\{1,1\}}\right)\delta_{il}\delta_{jk}\,,
\end{align}
and
\begin{align}
\label{eq:eedij}
\vev{\epsilon_{ij}\epsilon_{kl}^\dagger}=&V_{\{1,-1\}}\delta_{ij}\delta_{kl}+\frac{1}{N}(1-V_{\{1,-1\}})\delta_{il}\delta_{jk}\,.
\end{align}
At third order, there are four structures, but by complex conjugation one can reduce this to two unique ones:
\begin{align}
\vev{\epsilon_{ij}\epsilon_{kl}\epsilon_{mn}}=&\frac{1}{6}\left(V_{\{3\}}+4V_{\{2,1\}}+V_{\{1,1,1\}}\right)\delta_{ij}\delta_{kl}\delta_{mn}+\frac{1}{6}\left(V_{\{3\}}-V_{\{1,1,1\}}\right)(\delta_{il}\delta_{jk}\delta_{mn}+\delta_{in}\delta_{jm}\delta_{kl}+\delta_{ij}\delta_{kn}\delta_{lm}) \nonumber\\
&+\frac{1}{6}\left(V_{\{3\}}-2V_{\{2,1\}}+V_{\{1,1,1\}}\right)(\delta_{in}\delta_{jk}\delta_{lm}+\delta_{il}\delta_{kn}\delta_{jm})\,,
\end{align}
\begin{align}
\vev{\epsilon_{ij}\epsilon_{kl}\epsilon^\dagger_{mn}}=&\frac{1}{2}\left(V_{\{2,-1\}}+V_{\{1,1,-1\}}\right)\delta_{ij}\delta_{kl}\delta_{mn}+\frac{1}{2}\left(V_{\{2,-1\}}-V_{\{1,1,-1\}}\right)\delta_{il}\delta_{jk}\delta_{mn} \nonumber\\
&+\left(\frac{N}{(N-1)(N+1)}V_{\{1\}}-\frac{1}{2(N+1)}V_{\{2,-1\}}-\frac{1}{2(N-1)}V_{\{1,1,-1\}}\right)(\delta_{in}\delta_{jm}\delta_{kl}+\delta_{ij}\delta_{kn}\delta_{lm}) \nonumber\\
&+\left(-\frac{1}{(N-1)(N+1)}V_{\{1\}}-\frac{1}{2(N+1)}V_{\{2,-1\}}+\frac{1}{2(N-1)}V_{\{1,1,-1\}}\right)(\delta_{in}\delta_{jk}\delta_{lm}+\delta_{il}\delta_{kn}\delta_{jm})\,.
\end{align}

\section{Group Properties}
\label{apx:groups}
For a given group, the general basis is overcomplete.  These leads to simplifications in our derivations for a given group.  Here we present the related characters for three groups of relative importance: $U(1),SU(2),SU(3)$. 

For $U(1)$, the resulting identities are 
\begin{align}
\label{eq:u1idents}
 \chi_{\{1\}}&=-\chi_{\{1,1,-1\}}\,,\quad\chi_{\{1,\pm1\}}=\chi_{\{2,\pm1\}}=\chi_{\{1,1,1\}}=0 \,.
\end{align}
For $SU(2)$, one finds that
\begin{align}
\label{eq:su2idents}
\chi_{\{1\}}&=\chi_{\{2,1\}}\, ,\quad \chi_{\{2\}}=\chi_{\{1,-1\}}\,,\quad
\chi_{\{1,1\}}=1\,,\quad\chi_{\{1,1,1\}}=\chi_{\{1,1,-1\}}=0\,,\quad 
\chi_{\{3\}}=\chi_{\{2,-1\}}\, ,
\end{align}
and for $SU(3)$, the set of dependent representations needed up to third order in the cumulant expansion are
\begin{align}
\label{eq:su3idents}
\chi_{\{1\}}&=\chi_{\{1,1\}}\, ,\quad \chi_{\{2\}}=\chi_{\{1,1,-1\}},\quad
\chi_{\{1,-1\}}=\chi_{\{2,1\}}\, ,\quad \chi_{\{1,1,1\}}=1\, .
\end{align}

Another important set of identities are those which relate products of $\Re\chi_r$ to sum of $\Re\chi_r$.  They are easily enough derived, but we display a few key ones here:
\begin{align}
  \label{eq:ident2} (\Re\chi_{\{1\}})^2&=\frac{1}{2}\Re(\chi_{\{2\}}+\chi_{\{1,1\}}+\chi_{\{1,-1\}}+1)\,,\\
 \Re\chi_{\{1\}}\Re\chi_{\{2\}}&=\frac{1}{2}\Re(\chi_{\{1\}}+\chi_{\{2,1\}}+\chi_{\{2,-1\}}+\chi_{\{3\}})\,,\\
\Re\chi_{\{1\}}\Re\chi_{\{1,1\}}&=\frac{1}{2}\Re(\chi_{\{1\}}+\chi_{\{1,1,1\}}+\chi_{\{1,1,-1\}}+\chi_{\{2,1\}})\,,\\
 \Re\chi_{\{1\}} \Re\chi_{\{1,-1\}}&=\Re( \chi_{\{1\}}+\chi_{\{2,-1\}}+\chi_{\{1,1,-1\}})\,.
\end{align}

Applying all the simplifications in Eq.~\eqref{eq:fullaction} for specific groups,  we write $S[u]$ for $U(1)$ and $SU(2)$ respectively. For $U(1)$:
\begin{align}
  S[u]=\sum_p&-(\beta_{\{1\}}-\beta_{\{1,1,-1\}})\Re\chi_{\{1\}}\upfun-\beta_{\{0\}}\nonumber\\&-\beta_{\{2\}}\Re\chi_{\{2\}}-\beta_{\{3\}}\Re\chi_{\{3\}}\upfun \,. \label{eq: u1_action}
\end{align}
For $SU(2)$:
\begin{align}
   S[u]=\sum_p&-\left(\beta_{\{1\}}+\beta_{\{2,1\}}\right)\frac{1}{2}\Re\chi_{\{1\}}\upfun - \left(\beta_{\{0\}}+\beta_{\{1,1\}}\right)\nonumber\\&-\left(\beta_{\{2\}}+\beta_{\{1,-1\}}\right)\frac{1}{3}\Re\chi_{\{2\}}\upfun \nonumber\\&-(\beta_{\{3\}}+\beta_{\{2,-1\}}) \frac{1}{4}\Re\chi_{\{3\}}\upfun \,.
\end{align}

\section{Derivation of the Decimated Action}
\label{apx:deriv}
In this appendix, we expand upon the derivation of the decimated action.  First, for the second-order term in Eq.~\eqref{eq:second_order}, there are three terms which we decomposed based on the number of links that the two plaquettes $p,q$ shared. For case $p=q$ reads:
 \begin{align}
 \label{eq:21p}
 \beta^2\vev{\suep^2} &= \frac{\beta^2}{N^2} \langle \Re\left(\tr\left(u_1\epsilon_1u_2\epsilon_2(u_3\epsilon_3)^\dag(u_4\epsilon_4)^\dag\right)\right)\Re\left(\tr\left( u_1\epsilon_1u_2\epsilon_2(u_3\epsilon_3)^\dag(u_4\epsilon_4)^\dag\right)\right) \rangle\notag\\
 &=\frac{\beta^2}{2N^2}\left(|V_{\{1,1\}}|^4\Re\chi_{\{1,1\}}+|V_{\{2\}}|^4\Re\chi_{\{2\}}+V_{\{1,-1\}}^4\chi_{\{1,-1\}}+1\right)\,,
\end{align}
where we have utilized Eqs.~(\ref{eq:eeij}) and (\ref{eq:eedij}) to contract the $u$'s after integration.
The second term of Eq.~\eqref{eq:second_order} is obtained from first order action of Eq.~\eqref{eq: first-order-presentation} which reads,
\begin{align}
  \beta^2&\vev{\suep}^2 =\left(\frac{1}N \beta^2V_{\{1\}}^4\Re\chi_{\{1\}}\right)^2= \frac{1}{2N^2}\beta^2V_{\{1\}}^8\left( \Re\chi_{\{2\}}\upfun+\Re\chi_{\{1,1\}}\upfun+\chi_{\{1,-1\}}\upfun+1\right),
  \end{align}
  where we have used Eq.~(\ref{eq:ident2}).
  
  For the third-order terms of Eq.~\eqref{eq:connected2}, as discussed we need only consider when the three plaquettes are identical.
This will be done term by term, where the first term is: 
\begin{align}
  \beta^3\vev{\suep^3} &= -\frac{\beta^3}{N^3}\langle \Re\tr(u_1\epsilon_1u_2\epsilon_2(u_3\epsilon_3)^\dag(u_4\epsilon_4)^\dag)\Re\tr (u_5\epsilon_5u_6\epsilon_6(u_7\epsilon_7)^\dag(u_8\epsilon_8)^\dag)\Re\tr (u_9\epsilon_9u_{10}\epsilon_{10}(u_{11}\epsilon_{11})^\dag(u_{12}\epsilon_{12})^\dag)
  \rangle\nonumber\\
  &=-\frac{\beta^3}{2N^3}\bigg( \frac{V_{\{3\}}^4}{2}\Re\chi_{\{3\}} + V_{\{2,1\}}^4\Re\chi_{\{2,1\}}+\frac{V_{\{1,1,1\}}^4}{2}\Re\chi_{\{1,1,1\}} + \frac{3V_{\{2,-1\}}^4}{2}\Re\chi_{\{2,-1\}} \nonumber \\
  &\qquad\qquad+ \frac{3V_{\{1,1,-1\}}^4}{2}\Re\chi_{\{1,1,-1\}}+3V_{\{1\}}^4\Re\chi_{\{1\}}\bigg).\label{eq:first-term-thrid-order}
\end{align}
For the mixed-order term in Eq.~\eqref{eq:connected2}:
\begin{align}
  -&3\beta^3\vev{\suep}\vev{\suep^2} = \frac{3\beta^3V_{\{1\}}^4}{2N^3}\Re\chi_{\{1\}}\upfun \big[V_{\{2\}}^4\Re\chi_{\{2\}}\upfun +V_{\{1,1\}}^4\Re\chi_{\{1,1\}}\upfun+V_{\{1,-1\}}^4\chi_{\{1,-1\}}\upfun +1\big] \nonumber\\
  &= \frac{3\beta^3 V_{\{1\}}^4}{4N^3}\big[(V_{\{1,1\}}^4+2 V_{\{1,-1\}}^4+V_{\{2\}}^4+2)\Re\chi_{\{1\}}\upfun +(V_{\{1,1\}}^4+2 V_{\{1,-1\}}^4)\Re\chi_{\{1,1,-1\}}\upfun\nonumber\\
  &\hspace{2.2cm}+(V_{\{1,1\}}^4+V_{\{2\}}^4)\Re\chi_{\{2,1\}}\upfun +V_{\{1,1\}}^4 \Re\chi_{\{1,1,1\}}\upfun +(2V_{\{1,-1\}}^4+V_{\{2\}}^4)\Re\chi_{\{2,-1\}}\upfun +V_{\{2\}}^4 \Re\chi_{\{3\}}\upfun \big]\, , \label{eq:second-term-thrid-order}
\end{align}
where the second line was simplified with the identities from Appendix~\ref{apx:groups}.
The final term in Eq.~\eqref{eq:connected2} follows from another identity:
\begin{align}
  2&\beta^3\vev{\suep}^3 = -2 \frac{\beta^3V_{\{1\}}^{12}}{N^3}(\Re\chi_{\{1\}})^3 \label{eq:third-term-thrid-order}\nonumber\\
  &\quad = -\frac{\beta^3V_{\{1\}}^{12}}{2N^3}\big(\Re\chi_{\{3\}}\upfun +2\Re\chi_{\{2,1\}}\upfun +\Re\chi_{\{1,1,1\}}\upfun+6\Re\chi_{\{1\}}\upfun +3\Re\chi_{\{2,-1\}}\upfun +3\Re\chi_{\{1,1,-1\}}\upfun \big). 
\end{align}

\end{widetext}

\end{document}